\documentclass[final,3p,times,twocolumn,longtitle]{elsarticle}

\usepackage{graphicx}
\usepackage{amssymb}
\usepackage{subfigure}
\usepackage{floatflt}
\usepackage{captcont}

\newcommand{\unit}[1]{\ensuremath{\, \textrm{#1}}}
\newcommand{\degC}{\ensuremath{^{\circ} \textrm{C}}}
\newcommand{\muunit}[1]{\ensuremath{\, \mu \textrm{#1}}}
       
\journal{Astroparticle Physics}

\begin{document}
\begin{frontmatter}
  \title{Background studies for acoustic neutrino detection at the South Pole}
\author[Madison]{R.~Abbasi} 
\author[Gent]{Y.~Abdou} 
\author[RiverFalls]{T.~Abu-Zayyad} 
\author[Christchurch]{J.~Adams} 
\author[Madison]{J.~A.~Aguilar} 
\author[Oxford]{M.~Ahlers} 
\author[Madison]{K.~Andeen} 
\author[Wuppertal]{J.~Auffenberg} 
\author[Bartol]{X.~Bai} 
\author[Madison]{M.~Baker} 
% ...
\address[Christchurch]{Dept.~of Physics and Astronomy, University of Canterbury, Private Bag 4800, Christchurch, New Zealand}
% ...
\address[Gent]{Dept.~of Physics and Astronomy, University of Gent, B-9000 Gent, Belgium}
% ...
\address[Madison]{Dept.~of Physics, University of Wisconsin, Madison, WI 53706, USA}
% ...
\address[Bartol]{Bartol Research Institute and Department of Physics and Astronomy, University of Delaware, Newark, DE 19716, USA}
\address[Oxford]{Dept.~of Physics, University of Oxford, 1 Keble Road, Oxford OX1 3NP, UK}
\address[RiverFalls]{Dept.~of Physics, University of Wisconsin, River Falls, WI 54022, USA}
% ...
\address[Wuppertal]{Dept.~of Physics, University of Wuppertal, D-42119 Wuppertal, Germany}
% ...
\cortext[cor1]{Corresponding authors}
\fntext[Bari]{also Universit\`a di Bari and Sezione INFN, Dipartimento di Fisica, I-70126, Bari, Italy}
\fntext[Goddard]{NASA Goddard Space Flight Center, Greenbelt, MD
  20771, USA}

\begin{abstract}
  The detection of acoustic signals from ultra-high energy neutrino
  interactions is a promising method to measure the flux of
  cosmogenic neutrinos expected on Earth.  The energy threshold for
  this process depends strongly on the absolute noise level in the
  target material.  The South Pole Acoustic Test Setup (SPATS),
  deployed in the upper part of four boreholes of the IceCube Neutrino
  Observatory, has monitored the noise in Antarctic ice at the
  geographic South Pole for more than two years down to $500 \unit{m}$
  depth.  The noise is very stable and Gaussian distributed.  Lacking
  an in-situ calibration up to now, laboratory measurements have been
  used to estimate the absolute noise level in the $10$ to $50
  \unit{kHz}$ frequency range to be smaller than $20 \unit{mPa}$.
  Using a threshold trigger, sensors of the South Pole Acoustic Test
  Setup registered acoustic events in the IceCube detector
  volume and its vicinity.  Acoustic signals from refreezing IceCube
  holes and from anthropogenic sources have been used to test the localization
  of acoustic events.
  An upper limit on the neutrino flux at energies $E_\nu > 10^{11}
  \unit{GeV}$ is derived from acoustic data taken over eight months.
\end{abstract}

\end{frontmatter}

\section{Introduction}
\label{sec:introduction}

During recent years it has been extensively studied, whether the
glacier ice at the South Pole is a suitable material for the detection
of cosmic neutrinos above $10^{18} \unit{eV}$ energy, using their
acoustic signals emitted in the $10$ to $50 \unit{kHz}$ region.  With
data from the South Pole Acoustic Test Setup \cite{SPATSTECH}, the
speed of sound has been measured down to $500 \unit{m}$ depth.  It was
found to be constant below $200 \unit{m}$, favorable for neutrino
detection \cite{Abbasi:2009si}.  The sound attenuation in the $200$ to
$500 \unit{m}$ depth region was measured to be much stronger than
expected \cite{SPATSATT}.  An important precondition for acoustic
neutrino detection is a precise knowledge of the steady and transient
noise levels on top of which a possible neutrino signal has to be
detected.

Long duration noise studies in deep water of the Mediterranean Sea
\cite{Aguilar:2010oq,Riccobene:2009zz} and Lake Baikal
\cite{Aynutdinov:2009bs} have shown strong variations with time due to
changing environmental conditions such as wind speed and rain.
Sound sources such as ships and animals (e.g. sperm
whales) contribute as well.  In quiet periods, however, noise levels
as low as a few mPa are reported for the frequency region of interest.

Acoustic noise at the South Pole is assumed to be low and stable
because none of the sources mentioned above are expected to
contribute.  Due to the depth dependence of the density, the
first $200 \unit{m}$ of firn ice acts as an acoustic filter.
Anthropogenic and environmental noise entering the ice from the
surface will be refracted back.

The glacier moves with a velocity of about $10 \unit{m}$
per year over the bedrock beneath. Whether this or other phenomena
give rise to ultra-sound of detectable amplitude is one subject of the
investigations presented in this paper.

In Sec.~\ref{sec:setup}, the detector geometry, calibration and the
various data taking modes will be described.  In
Sec.~\ref{sec:noisefloor}, the permanent noise conditions will be
discussed, followed by information about the detection of sound
signals produced in the ice by construction activities of the IceCube
experiment \cite{Achterberg:2006md} in Sec.~\ref{sec:transients}.
Finally, a limit for the flux of cosmic neutrinos will be
derived from acoustic data accumulated during eight months.

\section{Detector setup and data taking}
\label{sec:setup}

\subsection{Detector geometry}

To measure the acoustic properties of the ice at the geographic South
Pole, the South Pole Acoustic Test Setup (SPATS) \cite{SPATSTECH}, a
system of four instrumented vertical lines, called strings A, B, C,
and D, was installed in boreholes of the IceCube neutrino observatory
after deployment of the IceCube optical modules.  Each string holds
seven stages, referred to as stages 1 to 7 from top to bottom; each
stage is a combination of an acoustic sensor and a transmitter,
vertically separated by $\sim\,1 \unit{m}$.  The horizontal distances
between strings range from $125 \unit{m}$ (which is the spacing
between IceCube strings) to $543 \unit{m}$.  Vertically, the depth
range from $80$ to $500 \unit{m}$ is instrumented with increasing
spacing of sensors in the deeper ice to be able to sample the
transition from the firn\footnote{The transition region from a
  snow/air mixture at the surface to solid ice is called firn. It has
  a width of about $170 \unit{m}$ at South Pole \cite{Abbasi:2009si}.}
region to the bulk ice.  On strings A, B, and C the stages are located
at depths of $80$, $100$, $140$, $190$, $250$, $320$, and $400
\unit{m}$; string D is deployed deeper in the ice and the stages are
located at $140$, $190$, $250$, $320$, $400$, $430$, and $500
\unit{m}$ depth.

The SPATS sensors consist of a cylindrical stainless steel pressure
housing with a diameter of $10 \unit{cm}$ in which three piezoelectric
elements are mounted to the wall, separated by $120$ degrees to
provide full azimuthal coverage.  A three-stage low noise
pre-amplifier with a gain of $10^4$ is attached directly to each
piezoelectric element and the analog signal is transmitted to the
surface via a shielded twisted pair cable.  On string D at depths of
$190 \unit{m}$ and $430 \unit{m}$, an alternative type of sensor,
HADES (Hydrophone for Acoustic Detection at South Pole), was deployed.
For the HADES sensors, the piezoelectric element and pre-amplifier
have been cast in resin and mounted below the housing.  This allows us
to study acoustic signals with different systematics introduced by the
sensor.

The sensor channels are each identified by a label of the form
$X\textrm{S}y(z)$, where $X$ is the letter of the string, S refers to
a sensor, $y$ is the number of the stage, and $z$ the channel number
within the sensor (running from 0 to 2).

\subsection{Data acquisition} 
\label{sec:daq}

On the surface, a read-out box buried in the snow above each string
contains an industrial PC, called string-PC, that is used for
digitization, time stamping, and storage of the data.  Each string-PC
is connected by a symmetric DSL connection to the SPATS master-PC that
is housed in the IceCube laboratory at the South Pole station.  The
master-PC collects the data from all four string-PCs, distributes a
GPS timing signal to them, and prepares the data for transfer to
the northern hemisphere via satellite or for tape storage.

Since 28 August 2008, SPATS has been operating as a detector for
transient acoustic signals.  Out of each hour, $45$ minutes are used
for triggered data taking while in the remaining $15$ minutes
environmental monitoring and system health data are recorded,
including measurements of the noise floor.  Details on the DAQ system
can be found in Ref.~\cite{SPATSTECH}.  To monitor the noise floor,
every hour the three channels of each sensor simultaneously record
$100 \unit{ms}$ of untriggered data with a sampling rate of $200
\unit{kHz}$. In transient data taking mode, three sensor channels from
each string are used, located at three different depths.  The data
stream from each channel, digitized with a sampling rate of $200
\unit{kHz}$, is continuously monitored in the string-PC. A sensor
noise band is defined for each channel $i$ spanning symmetrically
around the channels noise mean value $\mu_i$ from $\mu_i - 5.2 \,
\sigma_i$ to $\mu_i + 5.2 \, \sigma_i$ as illustrated in
Fig.~\ref{fig:adc_dist}, where $\sigma_i$ is the width of the noise
distribution of the respective channel. If an ADC value $x_i$ outside
the noise band ($x_i > \mu_i + 5.2 \, \sigma_i$ or $x_i < \mu_i - 5.2
\, \sigma_i$ is measured, $5 \unit{ms}$ of data ($1001$ samples) are
written to disk, centered on the triggering sample. The threshold of
$5.2 \, \sigma_i$ was chosen to fit the amount of data into the
allocated daily satellite bandwidth of $150 \unit{MB}$.  The detector
has been operated in two different modes, hereafter called mode 1 and
mode 2.  In mode 1, the sensors located at $190$, $250$, and $320
\unit{m}$ depth were used for transient data taking.  In mode 2 a
deeper sensor configuration ($250$, $320$, $400 \unit{m}$) was chosen.

One transient data file per string is written every hour. Transient
data taking runs are successful in $93\%$ of all cases. The remaining
runs commonly fail due to DAQ problems. No correlation between failing
runs and large amplitude pulses, which would bias the analysis
described in this work, has been observed. During measurements with
the acoustic ``pinger'' \cite{Abbasi:2009si, SPATSATT} large amplitude
pulses have been recorded with a rate of up to $10 \unit{Hz}$ without
an increase in the occurrence of failed runs.

\begin{figure*}
  \centering
  \subfigure[]{
    \includegraphics[width=6.5cm]{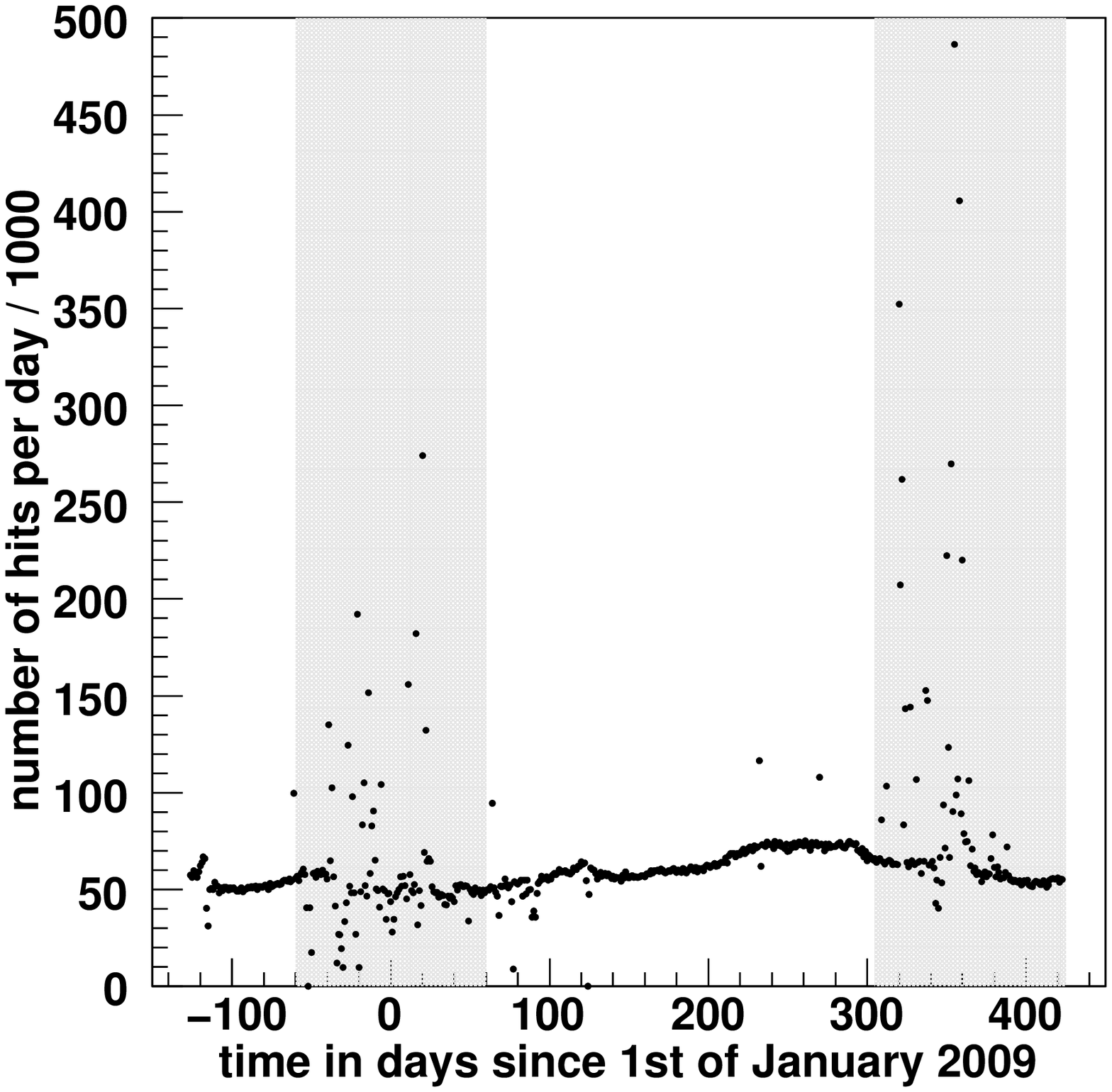}
    \label{fig:stat-hits}}
  \subfigure[]{
    \includegraphics[width=6.5cm]{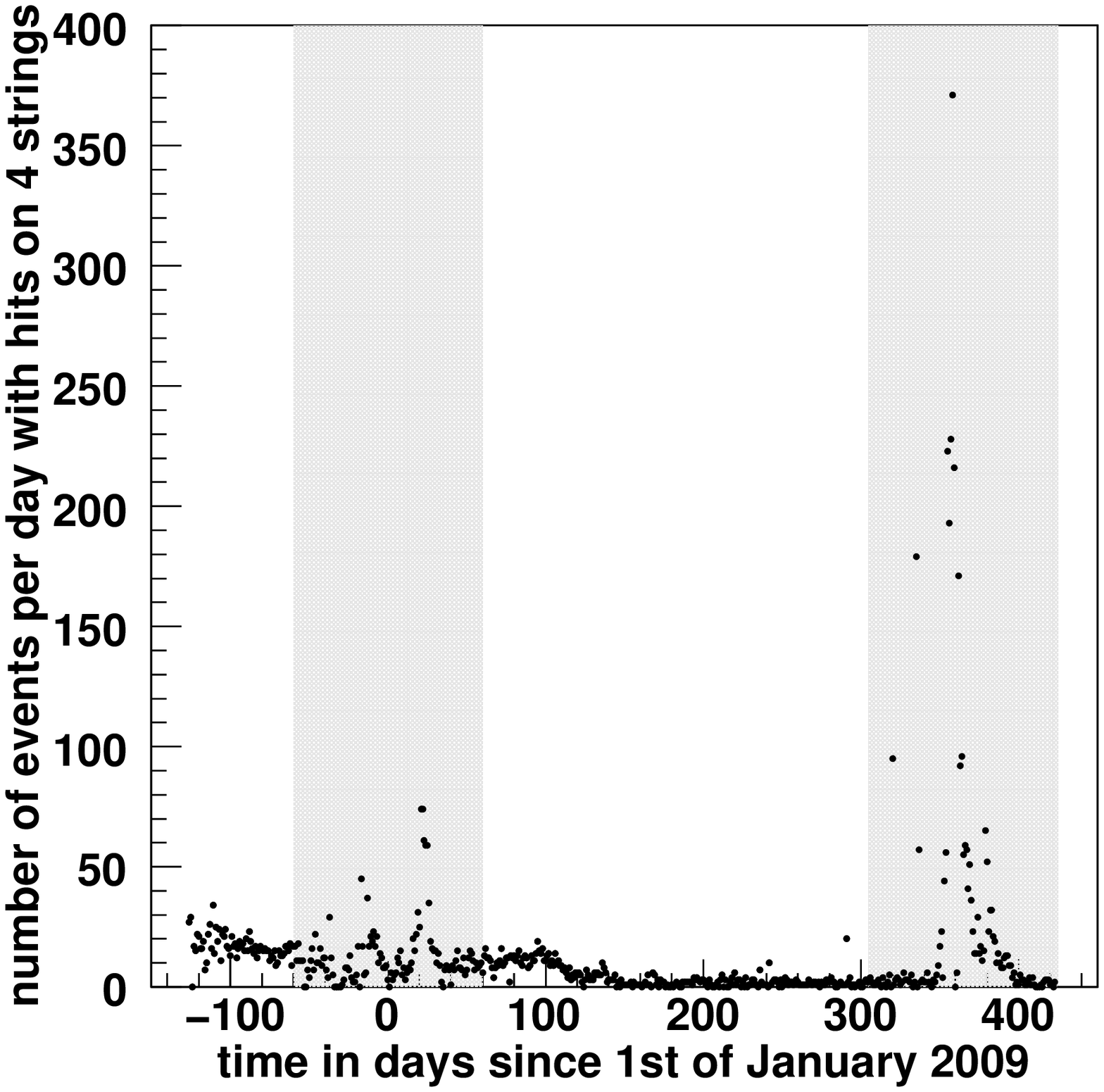}
    \label{fig:stat-4s}}
  \caption{(a) Number of acoustic hits per day and (b) acoustic events
    with hits at all $4$ strings since 1 January 2009. Drill periods
    are indicated in light gray.  For the definition of a ``hit'' and
    an ``event'' see Sec.~\ref{sec:transients}.}
\end{figure*}

As can be seen in Fig.~\ref{fig:stat-hits}, there are quiet periods
during the austral winter and more noisy periods during summer with
IceCube hole drilling and string deployment taking place.  We
distinguish between four periods of data taking; details can be found
in Tab.~\ref{tab:data_types}.  On average, about $65000$ triggers are
collected per day. The data are dominated by sensor noise,
i.e.~triggers caused by single ADC samples from the tails of the
Gaussian noise distribution. In the quiet periods, these account for
$99.99\%$ of all recorded waveforms, for which no coincident signals
in other sensor channels are observed.  Nearly all of these triggers
do not contain typical waveforms but have only a single spike larger
than the trigger threshold.

\begin{table*}
  \centering
  \caption{Characteristics for different data taking periods of
    transient noise triggers.}
  \label{tab:data_types}
  \begin{tabular}{|l|r|r|r|r|}
    \hline 
    Name & Quiet period 1 & Drill period 1 & Quiet period 2 & Drill
    period 2 \\ \hline \hline
    Start date & 28 Aug.~2008 & 1 Nov.~2008 & 1 Mar.~2009 & 1
    Nov.~2009 \\ \hline
    Duration (days) & $65$ & $120$ & $245$ & $120$ \\ \hline
    Available files & $5820$ & $9845$ & $22664$ & $10567$ \\
    \hline
    Avail./total & $0.93$ & $0.85$ & $0.96$ & $0.92$ \\ \hline 
    Detector mode & 1 & 1 & 2 & 2 \\ \hline
  \end{tabular} 
\end{table*}

\subsection{Sensor calibration}
\label{sec:calibration}

The sensitivity\footnote{Sensitivity is defined as the ratio $V / p$
  of the voltage $V$ induced at the output of an acoustic sensor,
  given a pressure amplitude $p$, with units of $\mathrm{V} \,
  \mathrm{Pa}^{-1}$.}  and equivalent self-noise\footnote{Equivalent
  self-noise is defined as the electronic self-noise of the device
  converted into an equivalent sound pressure level using the sensor's
  sensitivity.} of all SPATS sensors were measured in the laboratory
prior to deployment using the comparison method.  The mean sensitivity
(including the pre-amplifier) averaged over all sensor channels and
all frequencies in the relevant frequency band from $10$ to $50
\unit{kHz}$ is $2.8 \pm 0.8 \unit{V} \unit{Pa}^{-1}$ (equal to $-111
\, \textrm{dB re. } 1 \unit{V} \muunit{Pa}^{-1}$), where the error
indicates the spread between the different sensors.  The equivalent
self-noise in the same frequency band, averaged over all SPATS
sensors, is $7 \unit{mPa}$.  Details of the measurement are presented
in Ref.~\cite{SPATSTECH}.  The calibration was performed in water at a
temperature of $0 \degC$ and at normal pressure.  It is not clear a
priori, whether the sensitivity of the sensors remains unchanged when
deployed deep in Antarctic ice at a temperature of $-50 \degC$ and an
increased but not well known static ambient pressure.  The water in
the IceCube boreholes starts to freeze from the top and from the sides
so that cavities are formed in which the static pressure will increase
above the pressure of the water column at sensor depth.  After the
hole is completely frozen, relaxation of the freshly frozen ``hole
ice'' to the surrounding bulk ice occurs on an unknown time scale.
Furthermore, the sensitivity of the sensor can be influenced by the
different acoustic coupling, determined by the different acoustic
impedance matching between the water-sensor interface and the
ice-sensor interface, by the unknown horizontal position of the sensor
within the IceCube hole, and by possible shadowing of parts of the
sensor by the IceCube main cable running down the hole parallel to the
sensors.

Due to the absence of standardized acoustic sources for ice, it is not
possible to recalibrate the sensors after deployment. Also, it is
beyond the scope of the SPATS project to reproduce the combined
influence of very low temperatures, high static pressure and ice in a
laboratory environment.  However, we can study the influence of these
aspects on the sensitivity by separate investigations. In the
following discussion, we will assume that the influences of the
environmental parameters are uncorrelated and can thus be used to
estimate the in-situ sensor sensitivity from the results of different
laboratory experiments. A typical SPATS sensor was calibrated in air
at different temperatures.  A linear increase of its sensitivity with
decreasing temperature was measured.  The sensitivity increases by a
factor of $1.5 \pm 0.2$ from $0 \degC$ to $-50 \degC$
\cite{SPATSTECH}.  The same sensor was calibrated at room temperature
in a water filled pressure vessel at different static pressures from
$1 \unit{bar}$ to $100 \unit{bar}$.  No systematic change in the
sensitivity with pressure was observed; the sensitivity was found to
be stable within $30\%$ \cite{SPATSTECH}.  The effect of the acoustic
coupling of the sensor to the ice, which can lead to a flattening of
the frequency response function and a reduction of the sensitivity due
to a damping of mechanical resonances of the sensor housing, is not
yet understood and will be studied in an ice block in the laboratory
using the reciprocity calibration method (for a discussion of the
reciprocity calibration method, see e.g.~\cite{Urick:1983fk}).

Since the correlation between the different environmental parameters
is unknown when the sensor is exposed to a combination of all of them
in the deep Antarctic ice, the determination of the absolute noise
level is currently only roughly possible. Multiplying the sensor's
sensitivity changes due to temperature ($\times 1.5$) and due to
pressure ($\times 1$) and adding the uncertainties in quadrature, we
find that the sensitivity of the sensor in the deep ice is increased
by a factor of $1.5 \pm 0.4$ as compared to the value obtained in the
laboratory, resulting in a mean sensitivity of $4.2 \pm 1.6 \unit{V}
\unit{Pa}^{-1}$.

\section{Properties of the noise floor}
\label{sec:noisefloor}

\subsection{General}

We observe a Gaussian distribution of ADC values for each sensor
channel. Thus the noise can be characterized by two parameters: a
mean value and a standard deviation.  The mean value depends on an
instrumental DC offset in the different channels and is always close
to zero. The standard deviation is a measure for the noise level in
the sensor, which is a superposition of sensor electronic self-noise,
electromagnetic interference picked up on the signal cable from the
sensor to the surface\footnote{Electromagnetic interference is
  expected to be small since the signal is transmitted differentially
  from the sensor to the ADC on the surface.}, and possible acoustic
noise contributions from the surrounding ice.

\begin{figure}
  \centering
  \includegraphics[width=8.0cm]{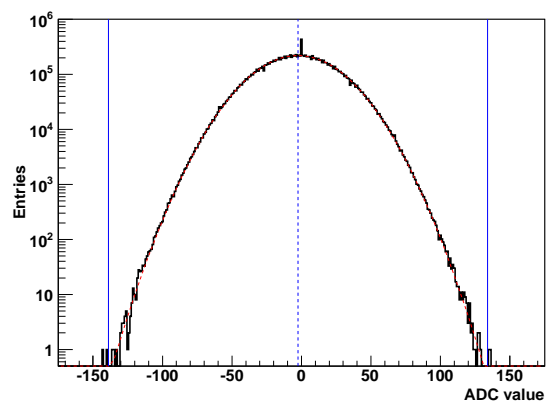}
  \caption{\label{fig:adc_dist}Distribution of ADC values from all
    noise data recorded for channel CS7(2) in July 2009 ($73.1
    \unit{s}$ of data in total). The red dashed line is a Gaussian fit
    to the data; the vertical lines indicate the mean value $\mu$
    (dashed) and the trigger thresholds at $\mu \pm 5.2 \,
    \sigma$. The peak at ADC value $0$ is an understood feature of the
    ADC.}
\end{figure}

The distribution of ADC values from $73.1 \unit{s}$ of recorded noise
data is shown in Fig.~\ref{fig:adc_dist}. The dynamic range of the
$12$-bit ADC is $\pm 5 \unit{V}$, corresponding to $2.4 \unit{mV}$ per
ADC count. The data are perfectly described by a Gaussian. The mean
value and trigger thresholds for transient data taking are indicated
in the graph. The four samples outside the noise band match very well
the expectation from the average SPATS trigger rate of $4.6$ triggers
during these $73.1 \unit{s}$.

\subsection{Stability}

\begin{figure}
  \centering
  \includegraphics[width=8.0cm]{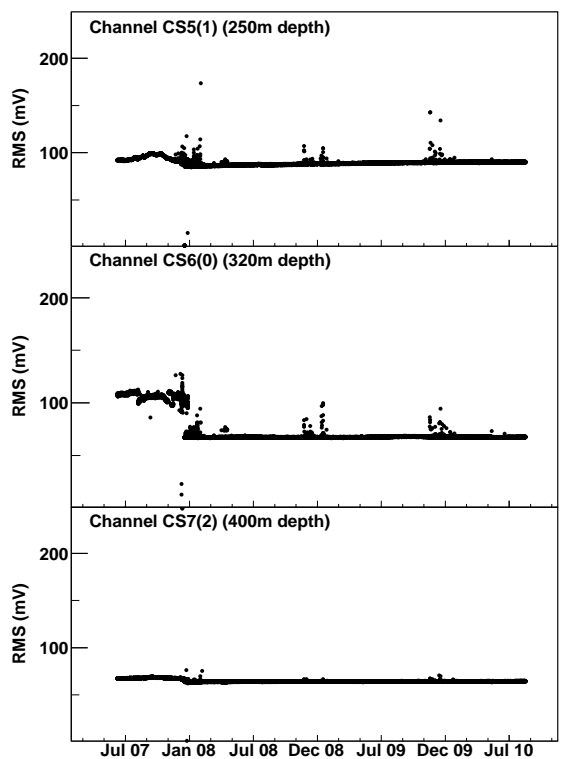}
  \caption{\label{fig:noiseRMS_vs_time}Noise RMS (from zero to $100
    \unit{kHz}$) as a function of time for three sensors on string C
    participating in transient data taking. The time window spans all
    data from deployment till today.}
\end{figure}

We have monitored the noise level in all sensor channels for more than
three years, beginning with the deployment of the first SPATS sensors
in January 2007.  Figure~\ref{fig:noiseRMS_vs_time} shows the RMS of
the noise as a function of time for three typical sensor channels on
string C that participate in the transient data taking.  All available
data from deployment until autumn 2010 is shown. It can be seen that,
apart from some short-time excesses that will be discussed below, the
noise level is very stable, the typical fluctuations being
$\sigma_{\mathrm{RMS}} / \langle \mathrm{RMS} \rangle < 10^{-2}$.

\begin{figure}
  \centering
  \includegraphics[width=8.0cm]{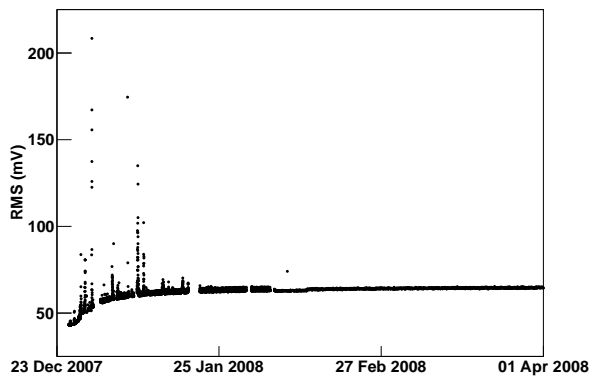}
  \caption{\label{fig:noiseRMS_vs_time_freeze}Noise RMS as a function
    of time during freeze in for sensor channel DS7(2) deployed at
    $500 \unit{m}$ depth.}
\end{figure}

In 2007, the first year of SPATS operation, we measured a higher and
less stable noise level.  During this period, the sensors were powered
on only for data taking and powered off afterwards which causes them
to heat up during the measurement and change their self-noise
characteristics.  Since December 2007, all sensors are powered on
continuously and are in thermal equilibrium with the surrounding ice.
Noise studies following a power outage show that it takes several
hours for a sensor to reach thermal equilibrium after it is powered
on.  The short-term noise excesses, which occur only in the Austral
summer seasons, can be correlated to IceCube deep-ice drilling
activity.  The visible spikes correspond to the holes drilled closest
to the SPATS array.  Due to technical reasons, data on the noise level
during the freeze in of the sensors after deployment is only available
for string D which was installed one year after the other three
strings.  It is shown for one sensor channel in
Fig.~\ref{fig:noiseRMS_vs_time_freeze}.

We observe an increase of the noise level after the deployment of the
sensor on 24 December 2007 that lasts for about three weeks, after
which the noise level became stable.  On top of that, excesses
correlated with IceCube deep-ice drilling can be seen.  We interpret
the rise of the noise level as a combination of the increase of
sensitivity with decreasing temperature
(cf.~Sec.~\ref{sec:calibration}) and an improved acoustic coupling of
the sensor to the bulk ice.

\subsection{Determination of the absolute noise level}
\label{sec:absolute-noise}

To determine the absolute noise level from the data, i.e.~the sound
pressure incident on the sensor, the sensitivity needs to be known. In
general, the sensitivity will be a function of direction and
frequency.  In the absence of a specific noise source model, an
equivalent acoustic power at the position of the piezoelectric element
is derived assuming isotropic noise. The power is then translated into
an effective pressure amplitude.

\begin{figure}
  \centering
  \includegraphics[width=8cm]{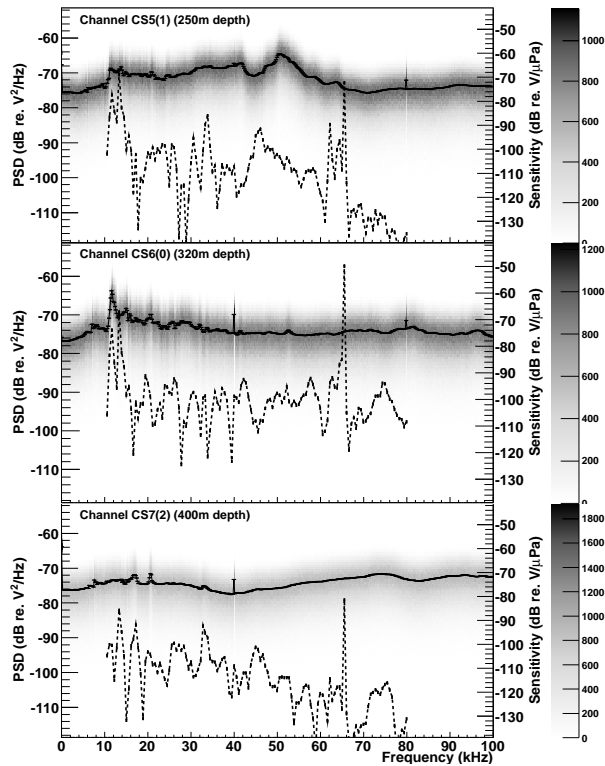}
  \caption{\label{fig:psd}Average power spectral density (PSD, shown
    as solid lines) and distribution for each frequency bin (gray
    scales) for three sensors on string C participating in transient
    data taking. All noise data recorded in July 2010 are shown. For
    comparison, the sensitivity of the sensor channels measured in the
    laboratory prior to deployment is shown as dashed lines.}
\end{figure}

In Fig.~\ref{fig:psd}, we show the voltage power spectral density (PSD)
distribution for three sensor channels participating in transient data
taking. The plot was obtained as follows: all noise floor data recorded
in July 2010 were Fourier transformed in sets of $1000$ samples each
(frequency resolution $\Delta f = 200 \unit{Hz}$) and the corresponding
PSD values were filled into a two dimensional histogram.  The gray scale
is a measure for the probability of occurrence of a certain PSD value at
a given frequency.  The solid lines represent the mean values,
calculated on a linear PSD scale, in each frequency bin.  The error of
the mean is also indicated, but too small to be visible.

\begin{figure}
  \centering
  \includegraphics[width=8.0cm]{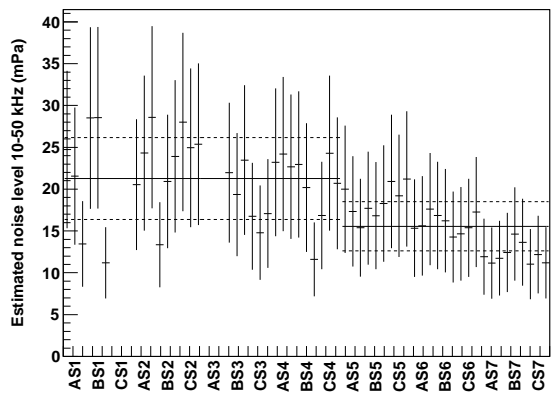}
  \caption{\label{fig:noise-level}Estimated absolute noise level
    integrated from $10$ to $50 \unit{kHz}$ for all SPATS
    channels. The error bars indicate the uncertainty on the
    sensitivity of the sensor channels. Sensors above $200 \unit{m}$
    depth (stages 1 to 4) and below $200 \unit{m}$ (stages 5 to 7) are
    treated separately (see text for details). Sensors AS3 and CS1 are
    broken and no data are available. The solid lines indicate the
    mean values and the dashed lines indicate the $1 \sigma$ spread of
    the data points. An average equivalent self-noise of $7
    \unit{mPa}$ can be subtracted quadratically for all sensors.}
\end{figure}

Figure~\ref{fig:psd} demonstrates that the spectral shapes differ
between all sensors. The sensitivity of the sensors as a function of
frequency, as measured in water in the laboratory prior to deployment,
is indicated as dashed lines. It is expected that the resonance
structure of the PSD is mainly governed by the mechanical response of
the sensor and not determined by the spectrum of the incident acoustic
background noise on the sensor. Especially the steel housing and its
coupling to the piezoelectric ceramic, which can be slightly different
for different sensors, should have a large effect. It can be seen in
Fig.~\ref{fig:psd} that peaks in the sensitivity are not reflected as
peaks in the PSD as would be expected for a smooth acoustic noise
spectrum.  This supports the assumption that the resonance behavior of
the sensor, and thus its sensitivity, is modified by the coupling of
the sensor housing to the ice.  Due to the suspected change in the
spectral sensitivity during freeze-in, we will \textit{not} calculate
the absolute noise level by dividing the PSD by the sensitivity to
determine the noise spectrum in units of pressure density and
integrate over the relevant frequency range.  This procedure would
introduce unknown errors by underestimating the contribution from
frequency regions with high sensitivity in the laboratory
calibration. Instead, we assume a single mean sensitivity for all
sensors, determined by averaging the laboratory sensitivity of each
sensor over the frequency range from $10$ to $50 \unit{kHz}$ and
subsequently averaging over all sensors and applying the correction
factor for temperature and pressure. This procedure yields a mean
sensitivity of $\langle S_{10-50} \rangle = 4.2 \pm 1.6 \unit{V}
\unit{Pa}^{-1}$ as discussed in Section~\ref{sec:calibration}.  We
determine the absolute noise level from each sensors voltage PSD
integrated from $10$ to $50 \unit{kHz}$, using the July 2010 data
presented in Fig.~\ref{fig:psd}. Attenuation losses in the cable of
$-0.6 \unit{dB} / 100 \unit{m}$ are corrected for. This assumes the
worst case scenario that all the measured noise is produced in the
sensor or is acoustic noise in the ice and no additional
electromagnetic noise is induced during transmission. Noise induced
further upstream in the DAQ chain would be over-corrected for cable
attenuation and result in an overestimation of the
noise. Figure~\ref{fig:noise-level} shows the resulting noise level
for all operative SPATS channels. We separate the sensors into two
groups: sensors above $200 \unit{m}$ depth and sensors below $200
\unit{m}$. The latter ones are used for the transient noise analysis
in the remainder of this work. For the shallow sensors, we calculate a
mean noise level of $21 \unit{mPa}$ with a $5 \unit{mPa}$ ($1 \sigma$)
spread between the data points. The average noise level in the deep
sensors is $(16 \pm 3) \unit{mPa}$.  This still includes the
contribution from electronic self-noise, that has been measured in the
laboratory prior to deployment to be $7 \unit{mPa}$ equivalent on
average.  Subtracting this contribution quadratically leads to an
estimated mean noise level in South Polar ice of $20 \unit{mPa}$
(shallow) and $14 \unit{mPa}$ (deep) integrated over the frequency
range relevant for acoustic neutrino detection of $10$ to $50
\unit{kHz}$.  Using the simulation described in Sec.~\ref{sec:limits}
and assuming an acoustic attenuation length of $300 \unit{m}$, a
pressure of $14 \unit{mPa}$ corresponds to the amplitude of the
acoustic signal generated by a neutrino of energy $10^{11} \unit{GeV}$
interacting in a distance of about $1000 \unit{m}$.

The origin and significance of the decrease of the noise level with
depth, that is visible in Fig.~\ref{fig:noise-level}, remains unclear.
One possible qualitative explanation for the observed depth dependence
is a contribution of noise generated on the surface. Due to the
gradient in the sound speed with depth \cite{Abbasi:2009si}, all noise
from the surface will be refracted back towards the surface, thus
shielding deeper regions from surface noise.

\section{Transient noise events}
\label{sec:transients}

A triggered waveform (``hit'')
contains $500$ samples measured before the trigger and $500$ samples
after the trigger, each separated by $5 \, \muunit{s}$ (see also
Sec. \ref{sec:daq}).  If the acoustic signal lasts longer than $2.5
\unit{ms}$, the following trigger is added to the hit under
consideration.  Hits from all strings are ordered in time offline and
merged into one file per day.  This file is processed through a
cluster algorithm to find all consecutive hits within $200 \unit{ms}$,
the time necessary for an acoustic signal to cross the SPATS array.
All hits per cluster are considered to form an acoustic event.  Events
with more than $5$ hits from at least $3$ strings are used to localize
their source position in the ice (see Section \ref{sec:reco}).  
The distribution in time of acoustic events with hits from all four strings
 is shown in Fig.~\ref{fig:stat-4s}.

\subsection{Source vertex reconstruction}
\label{sec:reco}

The acoustic event reconstruction algorithm is based on the solution
of the system of equations with $n=1,...,4$

\begin{equation}
  \hspace{-7mm}
  (x_n-x_0)^2+(y_n-y_0)^2+(z_n-z_0)^2=[v_s(t_n-t_0)]^2.
\end{equation}

Only four sensors with their signal arrival times and positions
$t_n,x_n,y_n,z_n$ are used in a single reconstruction.  The calculated
event vertex is located at the space time point $t_0,x_0,y_0,z_0$,
where the z-axis points vertically upward and $z=0$ corresponds to the
 ice surface.  The signal velocity in
ice is taken to be constant with $v_s = 3878 \unit{m/s}$
\cite{Abbasi:2009si}, and the propagation direction is assumed to be
straight.  The assumption of a constant speed of sound is only
suitable for events below a depth of around $200 \unit{m}$ and leads
to a spread of reconstructed event positions for shallower depths, as
one can see from simulations (Section \ref{sec:sim}).  Solving the
system of equations above provides an event vertex for a single four
sensor combination.  With twelve sensors in the used SPATS
configuration, statistical predictions can be made by using all
possible combinations $i=1,...,m$ of four sensors on four different
 strings per acoustic event.
In case of a noise hit in a sensor, the reconstruction algorithm for
this combination does not converge or the result lies far outside the
sensitive SPATS area.  The deviation from the mean vertex position of
all possible sensor combinations

\begin{equation}
  \bar{\vec{r}}=\frac{1}{m}\sum\limits_{i=1}^{m}\vec{r}_i,
\end{equation}

\noindent with $\vec{r}_i=(x_0^i,y_0^i,z_0^i)$, is used to improve
upon the background rejection. This is done by rejecting
reconstruction results for a single sensor combination if the distance
from the mean position in $x$,$y$ or $z$ is above $250 \unit{m}$.

\subsection{Acoustic event sources}
\label{sec:sources}

\begin{figure*}
\vspace{-2cm}
 \centering
  \subfigure[]{
    \includegraphics[width=12cm]{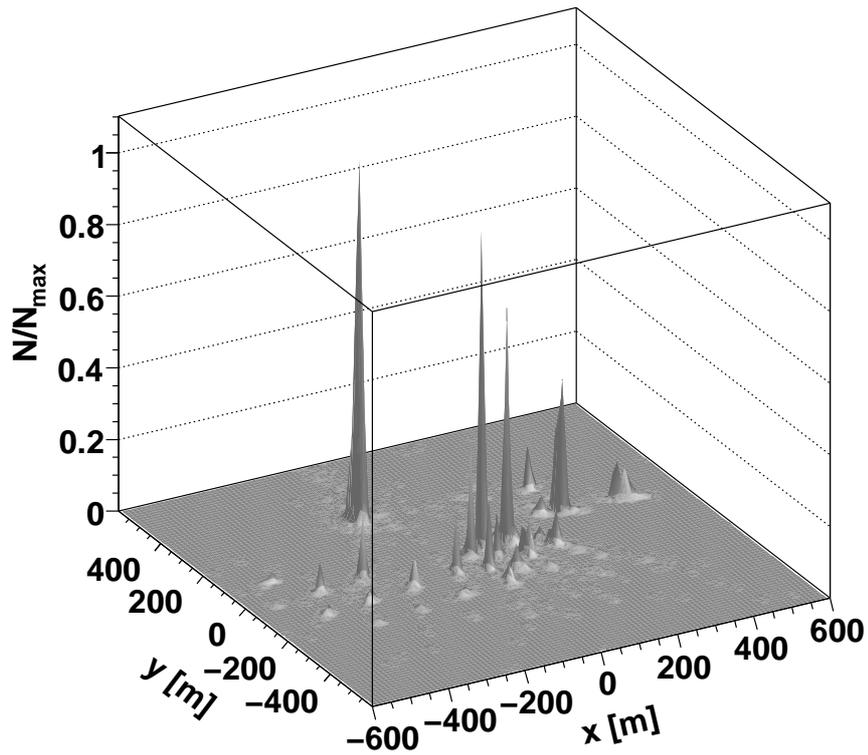}
    \label{fig:all-4s-XY-lego}}
  \subfigure[]{
    \hspace{-0.6cm}
    \includegraphics[width=8.5cm]{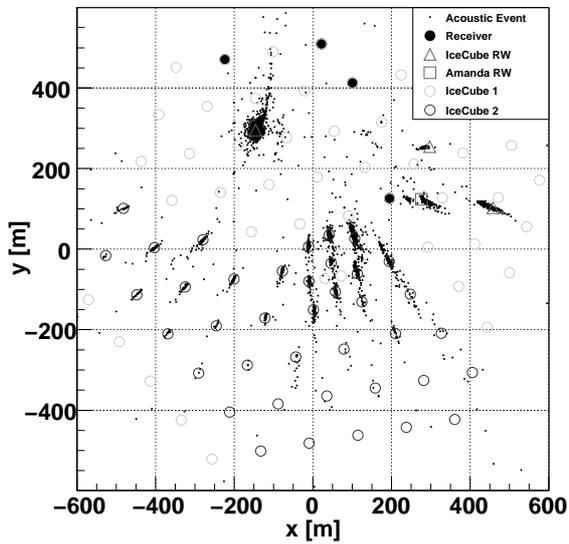}
    \label{fig:all-4s-XY}
    \hspace{-0.6cm}}
  \subfigure[]{
    \includegraphics[width=8.5cm]{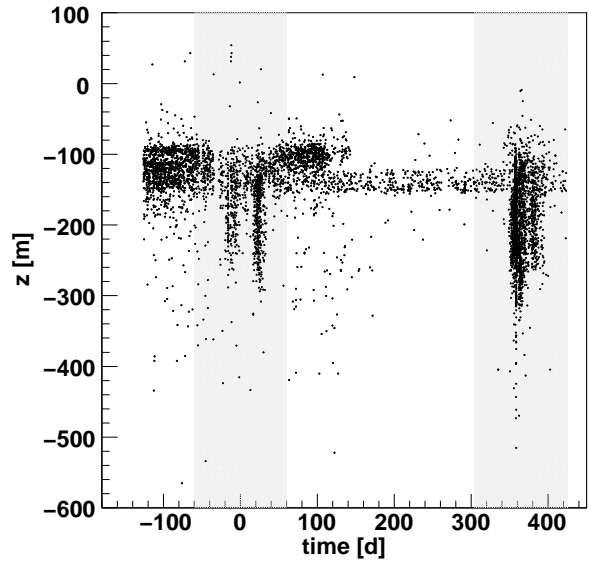}
    \label{fig:all-4s-ZT}}
  \caption{Shown are (a) the relative abundance of reconstructed acoustic 
    events in the
    horizontal plane of the IceCube coordinate system and (b) the
    actual vertex position of all transient events recorded since August
    2008.  The sources of transient noise are the Rodriguez Wells
    (RW), large caverns melted in the ice for water storage during
    IceCube drilling, and the refreezing IceCube holes.
    Dark gray circles (IceCube 2): positions of IceCube holes drilled in the
    period of transient data taking, light gray circles (IceCube 1):
    other IceCube holes, black filled circles: locations of SPATS strings, 
    triangles and square: location of RW. 
    The elongated structures are discussed in Sec.  \ref{sec:sim}.  
    In (c) the depth
    distribution of acoustic events versus time since August 2008 is
    shown. The zero at the time axis corresponds to Jan 1, 2009 and
    drill periods are indicated in light gray.}
\end{figure*}

In Fig.~\ref{fig:all-4s-XY-lego}, all reconstructed four-string events
are plotted according to their position in the IceCube coordinate
frame. Figure~\ref{fig:all-4s-XY} shows their location with respect to
the SPATS strings (large circles) and IceCube holes (small circles).
The triangles reflect the positions at which a Rodriguez-Well
(RW for short) \cite{RODWELL} is located (see Table
\ref{tab:RW-events}).  Such wells are used for the production and
cycling of water for the IceCube hot water drill system.  As can be
seen, the acoustic events are concentrated either at IceCube boreholes
or at Rodriguez-Well locations.  In Fig.~\ref{fig:all-4s-ZT}, the
event depth distribution versus time is shown. Almost no events are
located above 50 m depth.  In quiet periods, events are concentrated
between $80$ and $150 \unit{m}$. During drilling periods vertices are
still found down to $600 \unit{m}$.

\subsubsection{Events from IceCube holes}
\label{sec:holes}

Acoustic events were observed from nearly all IceCube holes drilled in
both seasons, when transient data taking was active.

\begin{figure*}
  \centering
  \subfigure[]{
    \includegraphics[width=6.5cm]{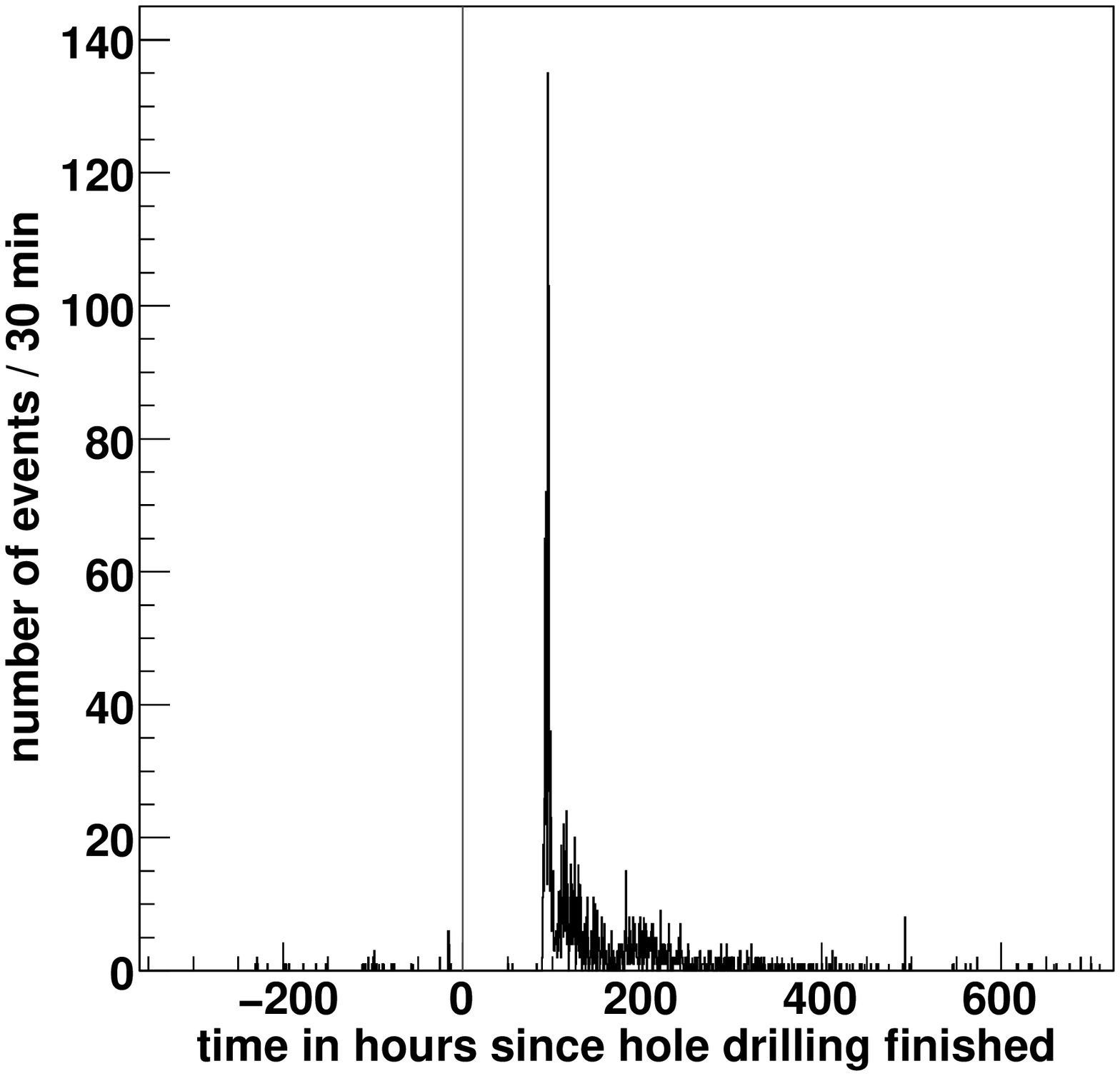}
    \label{fig:hole-81-T}}
  \subfigure[]{
    \includegraphics[width=6.5cm]{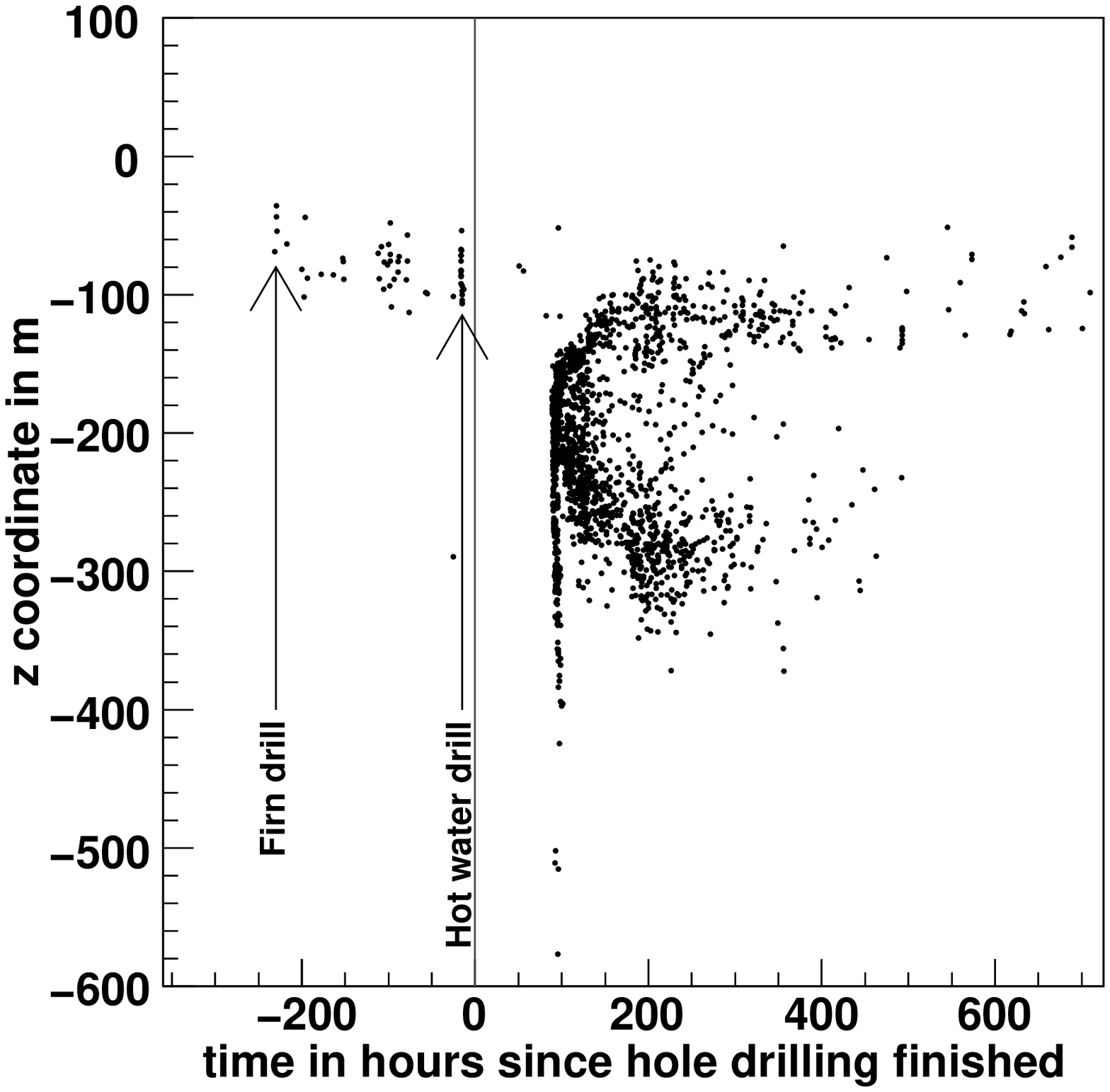}
    \label{fig:hole-81-TZ}}
  \caption{(a) Number of acoustic events per hour from hole 81 before
    and after drilling and (b) depth distribution versus time of
    acoustic events from hole 81. The vertical line indicates the end
    of hot water drilling for hole 81.}
\end{figure*}

\begin{figure*}
  \centering
  \subfigure[]{
    \includegraphics[width=6.5cm]{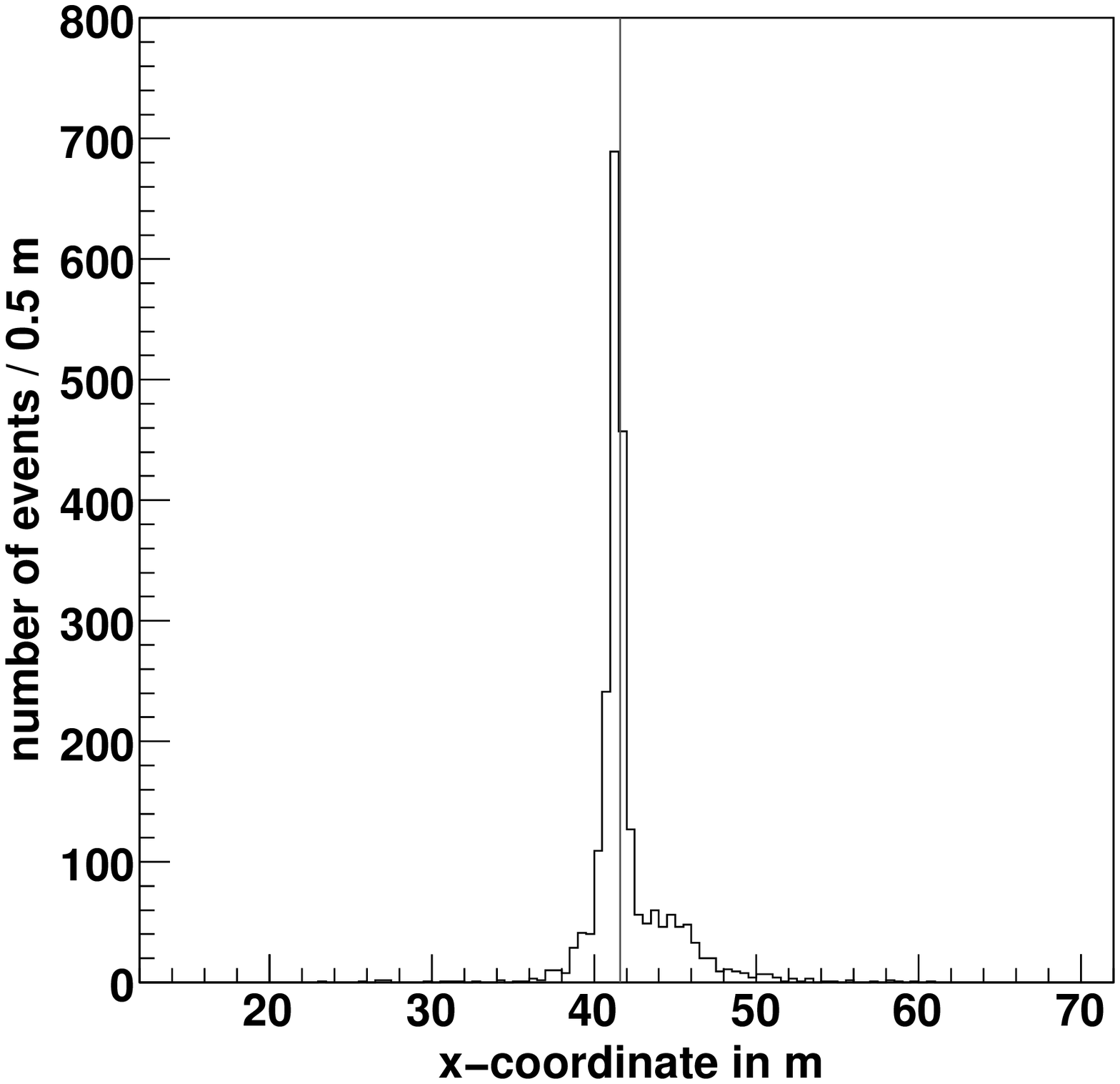}
    \label{fig:hole-81-X}}
  \subfigure[]{
    \includegraphics[width=6.5cm]{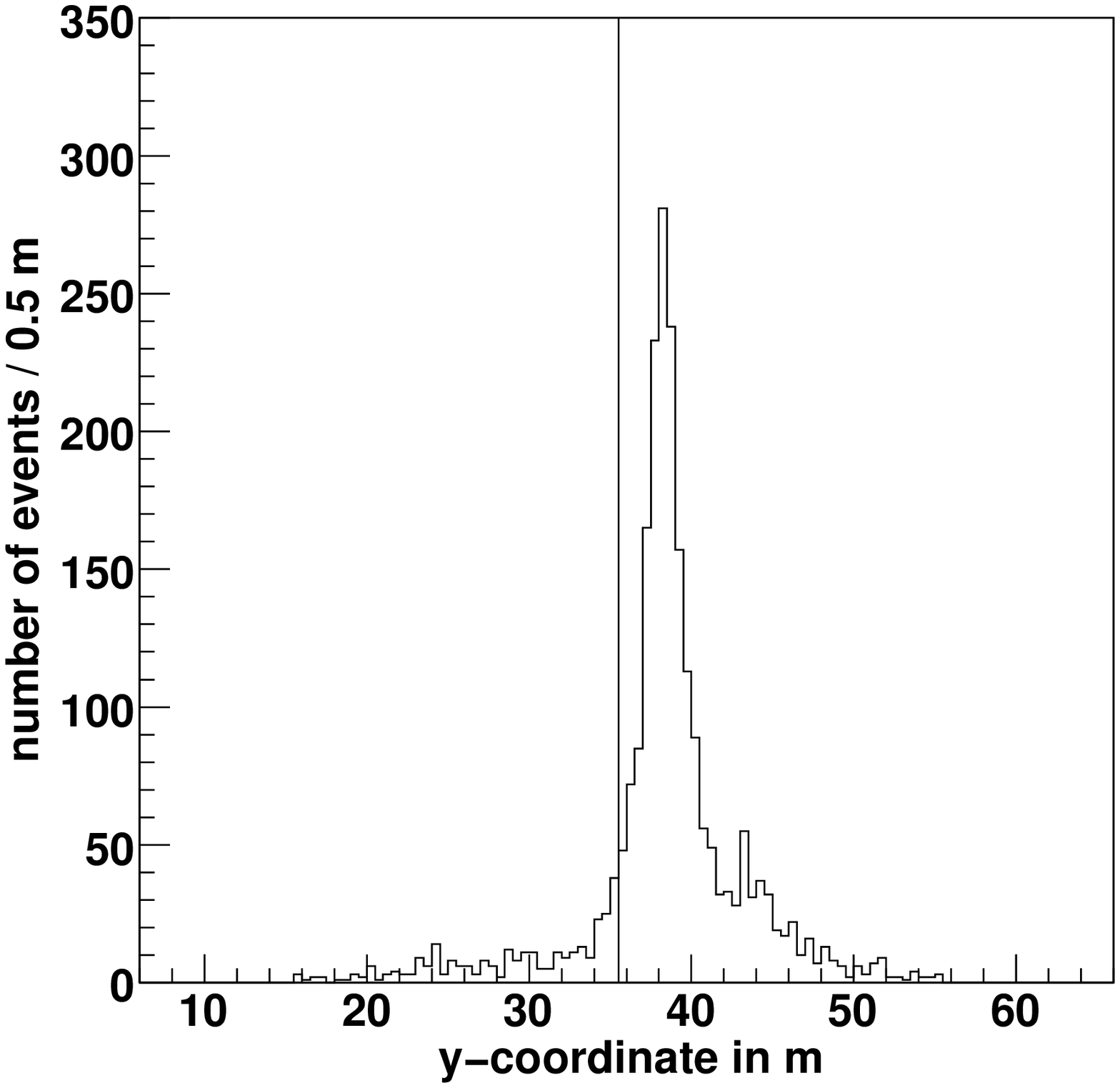}
    \label{fig:hole-81-Y}}
  \caption{(a) $x$-coordinate and (b) $y$-coordinate distribution of
    acoustic events from hole 81.  The vertical line shows the nominal
    position of hole 81.}
\end{figure*}

The statistics collected in the second season was much larger, due to
the fact that most of the 2009/10 holes were located close to the
center of the SPATS-detector.  The event distributions derived from
different holes are very similar. We investigate $2093$ events from
hole 81, the hole with the highest statistic, as an example.  Events
are observed for $20$ days in the hole region ($\pm 20 \unit{m}$ with
respect to the center of hole 81) during the periods of firn ice
drilling ($< 50 \unit{m}$ depth), bulk ice drilling ($50$ to $2500
\unit{m}$ depth) and refreezing, as can be seen in
Fig.~\ref{fig:hole-81-T}. Figure \ref{fig:hole-81-TZ} shows the depth
distribution of the events versus time. Before drilling, events are
observed at $40$ to $100 \unit{m}$ depth probably connected with noise
from the firn drill hole. During the procedure of hot water drilling a
few related events are found.  Strong sound production starts about
three days after drilling is finished, due to the refreezing
process. About $30\%$ of the registered events from this hole are
concentrated in two spots at $120 \unit{m}$ and $250 \unit{m}$ depth
but reach down to about $600 \unit{m}$. The reason is that the hole
does not refreeze homogeneously, but forms frozen ice plugs between
regions that are still filled with water.  The pressure produced in
this way may give rise to cracks near the ice water boundary which
would appear with sound in the $10$ to $100 \unit{kHz}$ frequency
region. Relaxation later continues within ``arms'' freezing towards
the hole surface and down to the lower ice plug (see
Fig. \ref{fig:hole-81-TZ}).  Besides providing information about the
refreezing process of water filled IceCube holes, one can also use the
corresponding acoustic events to understand the precision of the
vertex localization algorithm. In Fig.~\ref{fig:hole-81-X} and
Fig.~\ref{fig:hole-81-Y}, the reconstructed $x$ and $y$ event
positions are shown for hole 81 in the IceCube reference coordinate
system.  The average values including statistical uncertainties
for the $(x, y)$ position of hole 81 are determined to $(42.0 \pm 0.1)
\unit{m}$ in $x$ and $(38.5 \pm 0.1) \unit{m}$ in $y$.  The width of
the distributions is $2.6 \unit{m}$ and $5.0 \unit{m}$ respectively,
to be compared with a hole diameter of about $0.7 \unit{m}$. The
calculated values deviate from the actual hole positions at the
surface by $0.4 \unit{m}$ (in $x$) and $3.0 \unit{m}$ (in
$y$)\footnote{The actual positions of the sensors in the $x$-$y$ plane
  in the holes are known with a precision of $0.5 \unit{m}$ due to the
  hole width and inclinations versus depth from the drilling
  process.}. The possible reason for this deviation will be discussed
in the simulation section (Sec.~\ref{sec:sim}) below.

\subsubsection{Noise from Rodriguez-Wells}
\label{sec:rodwells}

When first acoustic events had been reconstructed during the first
period from August to November 2008 (quiet period), a strong
clustering in a certain region of the $x$-$y$ plane at about $(-150
\unit{m}, 300 \unit{m})$ became visible. It was found that this was
the position of the 2007/08 Rodriguez-Well used for the hot water
drilling system.

\begin{figure}
  \centering
  \includegraphics[width=7cm,height=10cm]{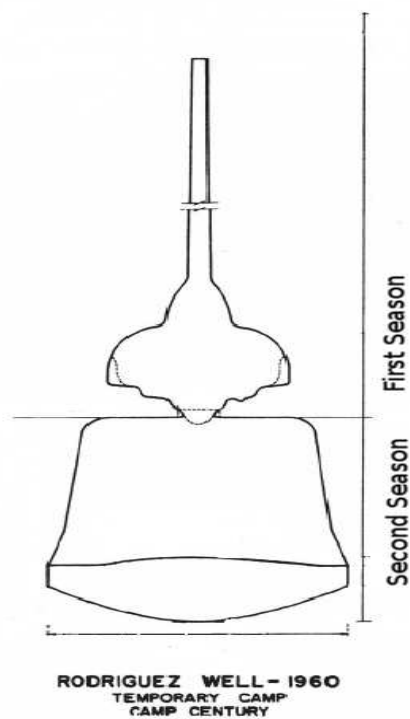}
  \caption{Section of the Camp Century well after a first and second
    season of operation (Schmidt and Rodriguez 1962) from
    \cite{RODWELL}.}
  \label{fig:Rodwell}
\end{figure}

This type of well has been introduced by Rodriguez and others in the
early 1960s \cite{RODWELL} to supply water from a glacier in
Greenland.  In Fig.~\ref{fig:Rodwell}, a sketch of the installation is
shown. Hot water cycled by a pump system is used to melt ice below the
firn layer at $60$ to $80 \unit{m}$ depth to maintain a fresh water
reservoir. An expanding cavity is formed with a diameter as large as
$15$ to $20 \unit{m}$. For IceCube and its predecessor AMANDA this
technique is used in connection with drilling at the South Pole since
mid 1990s. If the well is used a second time a year later, a second
cavern is formed at a deeper level.

\begin{figure*}
  \centering
  \subfigure[]{
    \includegraphics[width=6.5cm]{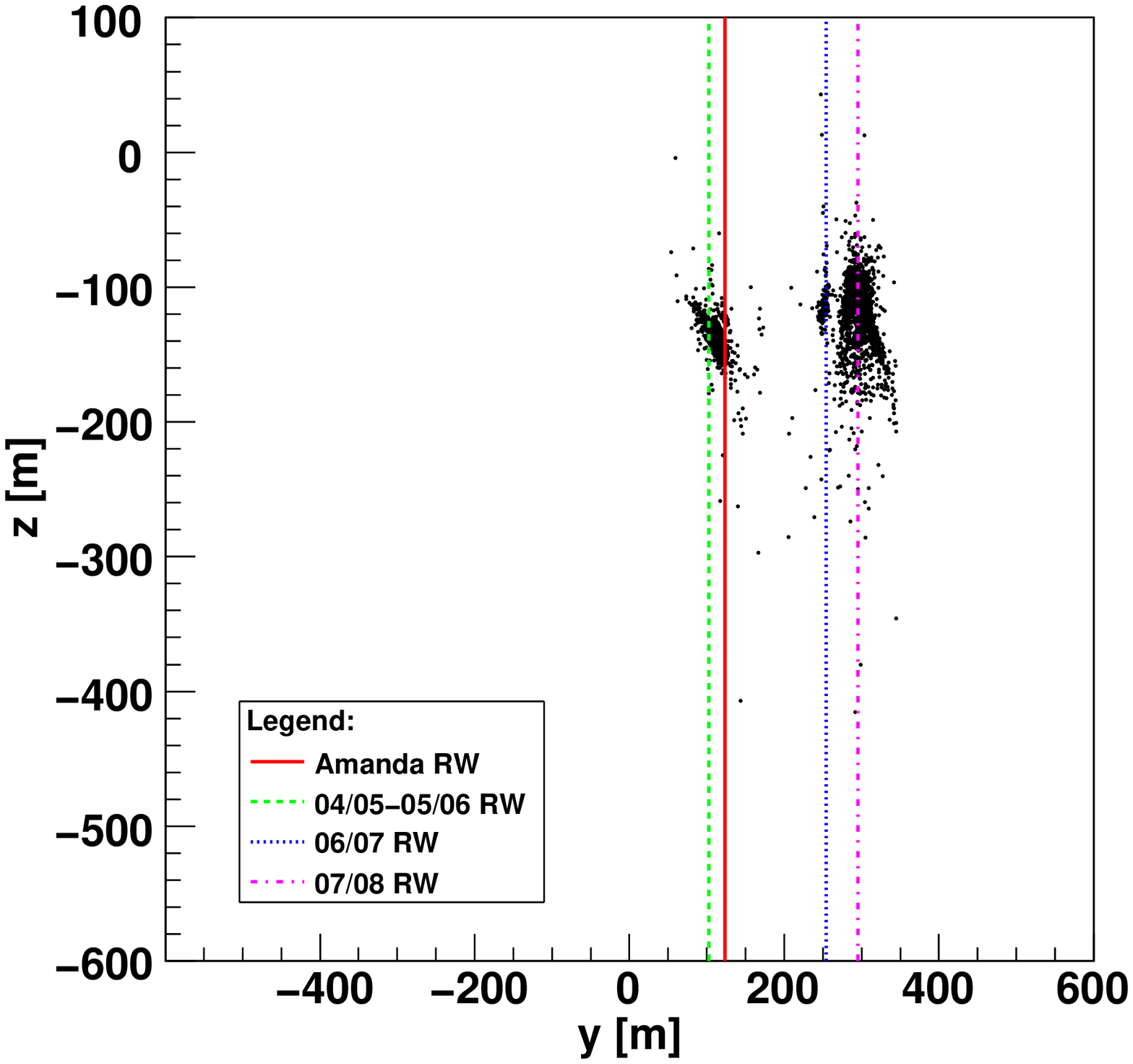}
    \label{fig:RW-yz}}
  \subfigure[]{
    \includegraphics[width=6.5cm]{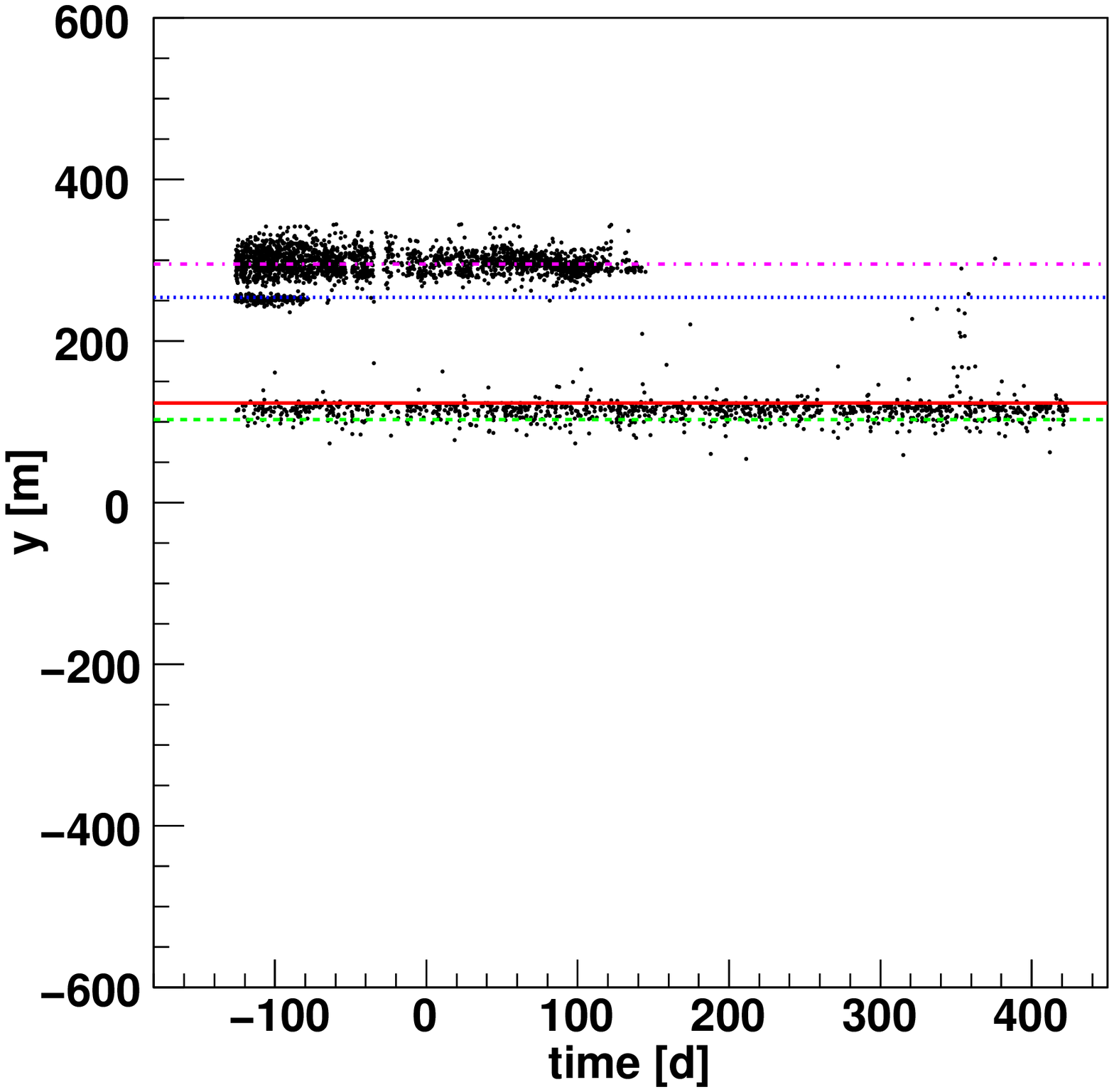}
    \label{fig:RW-yt}}
  \caption{Acoustic events distribution in (a) depth versus
    $y$-coordinate and (b) in $y$-coordinate versus time.  Lines:
    fitted positions of AMANDA-RW (solid red), 04/05-05/06 IceCube-RW
    (dashed green), 06/07 IceCube-RW (dotted blue), 07/08 IceCube-RW
    (dashed-dotted magenta). The time is given in days relative to
    Jan.~1, 2009.}
\end{figure*}

\begin{figure}
  \centering
  \includegraphics[width=8cm]{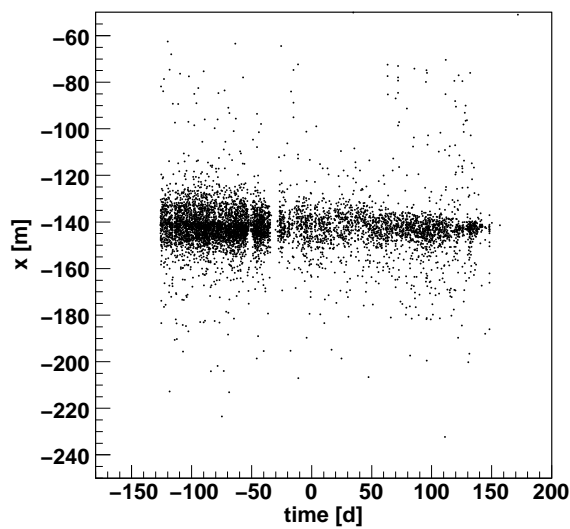}
  \caption{Re-freezing of Rodriguez-Well 2007/08 in the $x$-$t$
    plane. Acoustic signals are emitted from an decreasing area around
    the well core until they disappear in May 2009. The time is given in days
    relative to Jan. 1,2009.}
  \label{fig:RW08_xy_grey}
\end{figure}

Having identified acoustic events arising from the 2007/08
Rodriguez-Well, three other event clusters were found, two of them
could be attributed to other IceCube Rodriguez-Wells from 2006/07 and
2004 to 2006. The fourth event cluster turned out to be located at the
probable position of the last AMANDA Rodriguez-Well used in the final
two drilling seasons up to 2001. No documented coordinates could,
however, be found for that position. Available information about
acoustic event clusters connected with Rodriguez-Wells is summarized
in Table~\ref{tab:RW-events}. As can be seen from the table and from
Fig.~\ref{fig:RW-yz}, the acoustic events from the two Rodriguez-Wells
used only during one season are located at shallower depths than those
from Rodriguez-Wells used twice. This is in agreement with
expectations from the sketch in Fig.~\ref{fig:Rodwell}. The former
were seen to emit acoustic signals from regions of decreasing volume
around the well core and finally stopped, the older one in October
2008, the younger one in May 2009. As an example,
Fig.~\ref{fig:RW08_xy_grey} shows the time profile of refreezing for
Rodriguez-Well 2007/08, which was used only during one season. In
contrast to that, acoustic events are observed until today from the
six and ten years old deeper wells (see Fig.~\ref{fig:RW-yt}). The
mechanism of sound production in and around the Rodriguez-Well caverns
is still under debate in particular for the older wells.

\begin{table*}
   \centering
   \caption{Positions of acoustic events from different Rodriguez-Wells.
     The errors are given by the rms values of the event distribution and
     the asterisk in the last column marks the date when our analysis stopped.}
   \label{tab:RW-events}
   \begin{tabular}{|l|r|r|r|r|r|c|l|}
     \hline
     Name & $x_{nom} [m]$ & $y_{nom} [m]$ & $x_{fit} [m]$ & $y_{fit}
     [m]$ & $z_{fit} [m] $ & used & seen until \\ \hline \hline
     AMANDA & -- & -- & $276.2 \pm 0.4$ & $123.6 \pm 1.0$ & $-147.3 \pm
     1.1$ & $2 \unit{y}$ & Feb.~10$^{\ast}$ \\ \hline
     IC-RW 04-06 & $458.6$ & $102.7$ & $412.6 \pm 5.3$ & $124.0 \pm
     2.4$ & $-147 \pm 11$ & $2 \unit{y}$ & Feb.~10$^{\ast}$ \\ \hline
     IC-RW 06/07 & $297.2$ & $254.4$ & $279.6 \pm 0.4$ & $252.2 \pm
     1.0$ & $-114.2 \pm 0.7$ & $1 \unit{y}$ & Oct.~08 \\\hline
     IC-RW 07/08 & $-145.6$ & $295.6$ & $-138.6 \pm 0.4$ & $297.7 \pm
     0.6$ & $-118.3 \pm 1.0$ & $1 \unit{y}$ & May~09 \\\hline
   \end{tabular}
\end{table*}

\subsection{Acoustic event simulation}
\label{sec:sim}

A simple acoustic transient event simulation is done by calculating
the signal propagation times for the distance
$d_n=\sqrt{(x_n-x)^2+(y_n-y)^2+(z_n-z)^2}$ between source
(e.g.~IceCube hole at $(x, y, z)$) and sensors
$n=1,...,n_\mathrm{max}$ with $\Delta t_n=d_n/v_s$.  The signal is
transmitted from a random point inside a certain cylindrical volume
(radius $2 \unit{m}$, depth $2000 \unit{m}$) around the source.

\begin{figure}
  \centering
  \includegraphics[width=8.5cm]{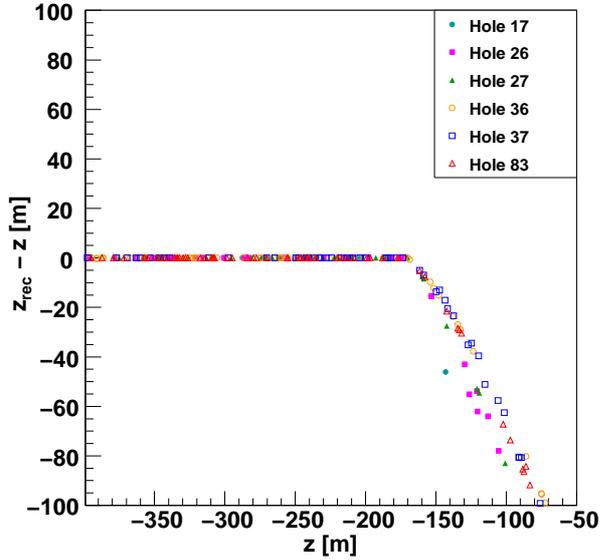}
  \caption{Difference between reconstructed event position
    and true event position for simulated events.
    Deviations seen in the upper region
    ($z>-170 \unit{m}$) are caused by the depth dependent sound
    speed.}
  \label{fig:sim-vel-att-drillphase-oldc-xz}
\end{figure}

Although knowing that the true IceCube hole diameter is about $0.7
\unit{m}$, we take into account the possibility that tension cracks
might appear outside the hole bounding surface, suggesting a larger
simulation radius. The reconstruction of events simulated with
constant speed of sound and without considering attenuation effects
implies an exact source localization, which is in contradiction to the
real data vertex results, where a specific spread of vertices around
the source (Fig.\ref{fig:all-4s-XY}) and a lack of events below and
above a certain depth is visible (Fig. \ref{fig:hole-81-TZ}). The
major reason for misreconstruction of events at shallow depth ($-200
\unit{m} < z < -1 \unit{m}$) is the depth dependence of the sound
speed \cite{Abbasi:2009si}, which is therefore included in the
simulation.  Above $174.8 \unit{m}$ and below $1 \unit{m}$ depth, the
parameterization
 
\begin{equation}
  \label{param}
  \hspace{-7mm} v_s=-(262.379+199.833
  ~\Bigl|\frac{z}{\mathrm{m}}\Bigl|^{\frac{1}{2}}-1213.08
  ~\Bigl|\frac{z}{\mathrm{m}}\Bigl|^{\frac{1}{3}}) \unit{m}
  \unit{s}^{-1}
\end{equation}

\noindent is used.  Below a constant sound speed value of $v_s=3878
\unit{m} \unit{s}^{-1}$ is assumed. Parameterization of
Eq. \ref{param} is obtained using a fit of in-ice sound speed data
points \cite{Weihaupt:1963,Abbasi:2009si}.  These conditions are taken into
 account by integrating along the path between source and sensor
using the parameterized value of $v_s$ at every point.  Due to the
absolute positions of the source and the sensors and their relative locations to
 each other, refraction effects are negligible as shown e.g. in Fig. 6 of
 reference \cite{Abbasi:2009si} and are thus neglected in this paper.  
Further improvements are achieved by using additional information on the
acoustic pressure wave attenuation in ice.  We apply the formula

\begin{equation}
  S_R^n=\left(\frac{S_0 \cdot d_0}{d_n}\right)
  e^{- \frac{d_n-d_0}{\lambda}}~,
\end{equation}

\noindent with the initial amplitude $S_0$ at the distance $d_0$
chosen to fit the real data and an attenuation length $\lambda = 300
\unit{m}$ as measured for South Pole ice with SPATS
\cite{SPATSATT}. $S_R^n$ is the corresponding signal amplitude at the
sensor $n$.  If the signal strength at a sensor is above $\sim 300
\unit{mV}$ ($5.2 \sigma$ above the noise level), a hit is triggered as
in real data.

Good agreement between reconstructed real and reconstructed simulated
events is obtained below 170 m depth, where the localisation precision in
 the z coordinate is 25 cm (see Fig.\ref{fig:sim-vel-att-drillphase-oldc-xz}). 
As expected, we observe a large influence of the
depth dependence of the sound speed on reconstructions in the upper region of
SPATS (between $0$ and $170 \unit{m}$ depth), as one can also see in
Fig. \ref{fig:sim-vel-att-drillphase-oldc-xz}.  The significant
deviation of the sound speed from the constant value used in the
reconstruction explains the spread of vertices seen in the real data
(see Fig. \ref{fig:all-4s-XY}), whereas the direction of this smearing
is caused by the detector geometry.  Due to the strong attenuation, it
is more difficult to observe deep events, which is well reproduced by
the simulation.

\section{Estimated neutrino flux limit}
\label{sec:limits}

\begin{figure*}
  \centering
  \subfigure[]{
    \includegraphics[width=6.5cm]{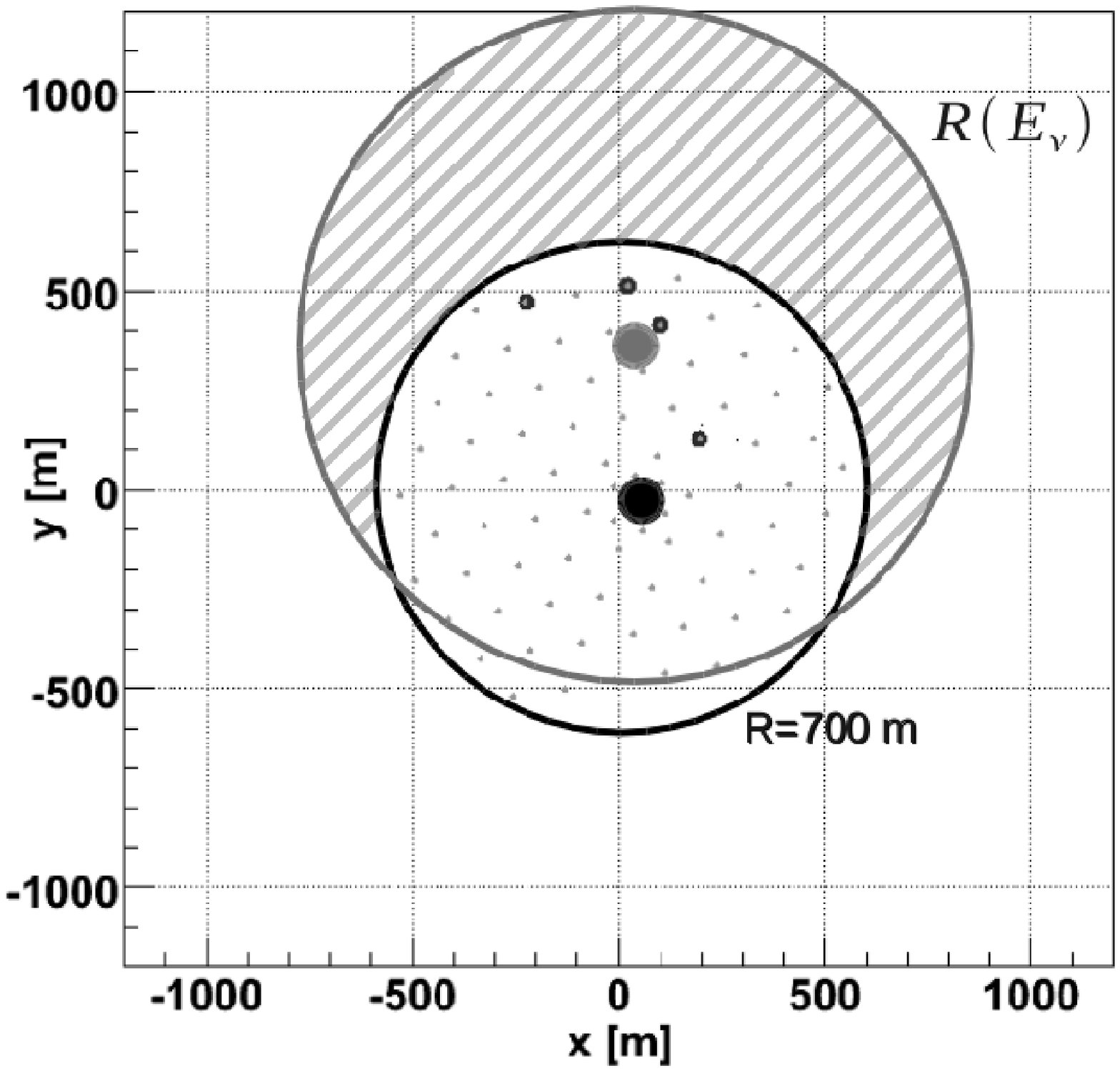}
    \label{fig:effvolume-geometry}}
  \subfigure[]{
    \includegraphics[width=6.5cm]{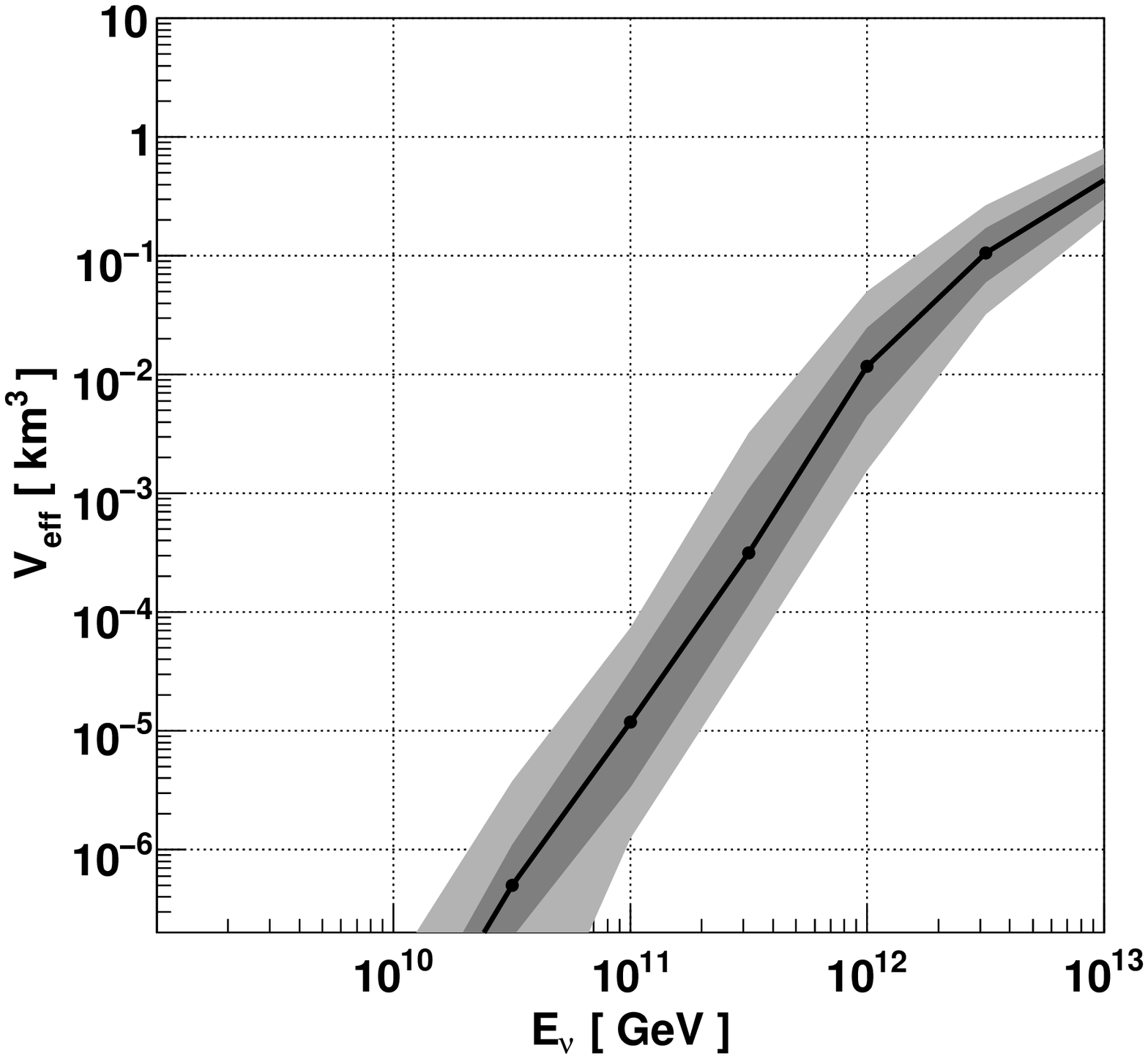}
    \label{fig:effvolume}}
  \subfigure[]{
    \includegraphics[width=10cm]{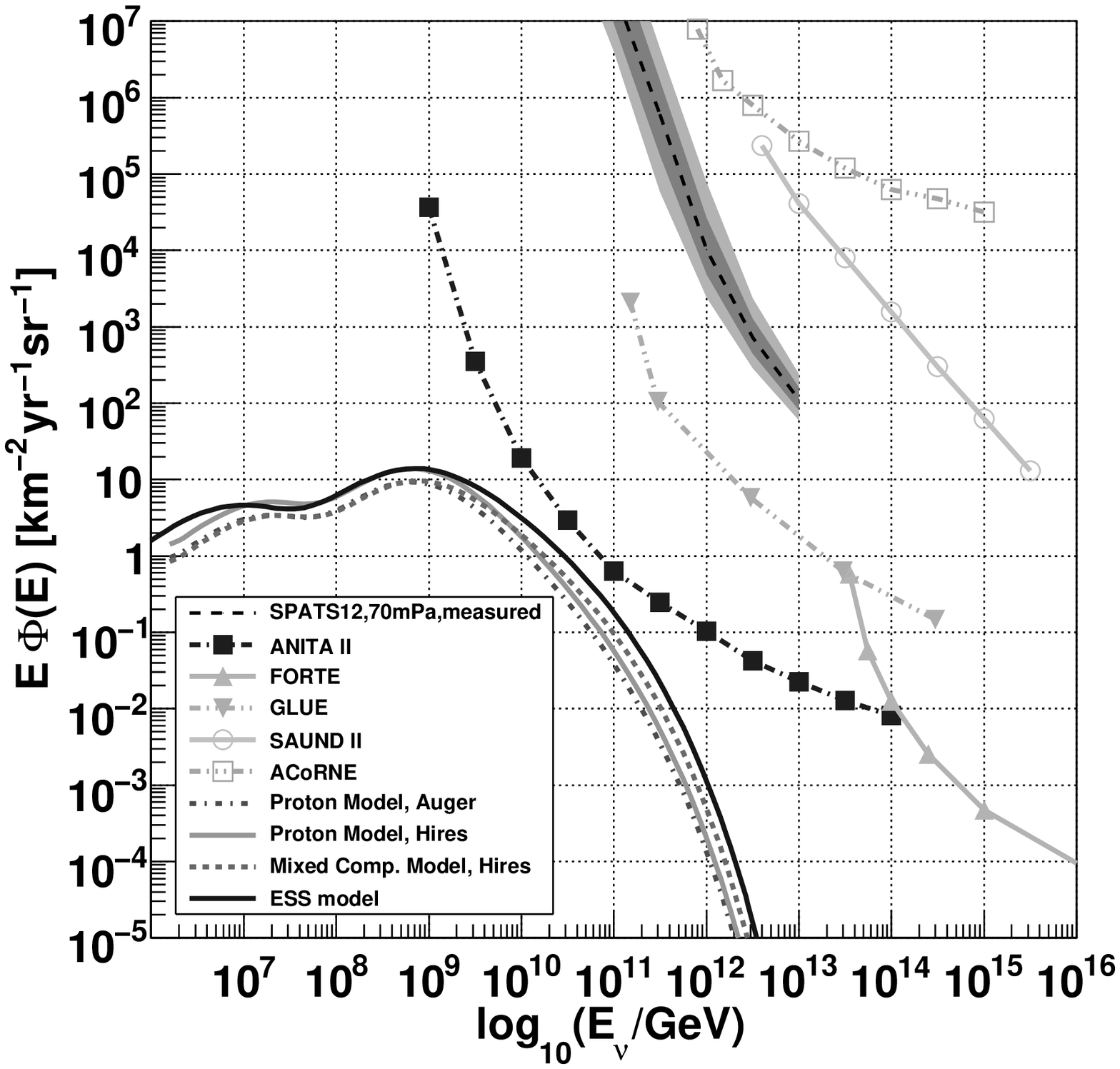}
    \label{fig:sensitivity}}
  \caption{(a) Shown are the IceCube center (large black dot) with the
    construction area (black circle) and the SPATS center (large grey
    dot) with the sensitive area for acoustic detection (grey circle).
    In addition, strings with acoustic sensors (medium size dots) and
    all IceCube strings (small dots) are drawn in.  To avoid events
    from IceCube construction, only the shaded area of the SPATS
    sensitive region is used in order to measure the neutrino flux
    limit.  (b) Corresponding effective volume and (c) the neutrino
    flux limit of the 2009 SPATS configuration ($70 \unit{mPa}$
    threshold, $\geq 5$ hits per event).  The dark gray band ($50$ to
    $100 \unit{mPa}$ threshold) around the effective volume and around
    the limit considers uncertainties in absolute noise. The even
    broader light gray band includes additional uncertainties due to
    the choice of different acoustic models.  Experimental limits on
    the flux of ultra high energy neutrinos are from ANITA II
    \cite{Gorham:2010kv}, FORTE \cite{Lehtinen:2003xv}, GLUE
    \cite{Gorham:2003da}, SAUND II \cite{Kurahashi:2010ei}, ACoRNE
    \cite{Bevan:2009zz}. For model reference see \cite{Engel:2001hd,Seckel}.}
\end{figure*}

Although SPATS has not been built to measure a relevant neutrino flux
limit, it is interesting to find out how sensitive a corresponding
measurement could be, using data from this setup.  In order to
determine the number of events not connected to IceCube construction
activities in the sensitive region of SPATS, we omit the area of
IceCube strings and the data from the drill periods.  The area taken
into account is indicated by the hatched area of
Fig. \ref{fig:effvolume-geometry}.  Furthermore, we look at depths
between $200$ and $1000 \unit{m}$, in the region of constant speed of
sound, to avoid the smearing effect in the reconstruction of acoustic
event locations described in Section \ref{sec:sim}.  In the $245$ days
of transient data taking (quiet period 2) we found no events in the
defined region.  This observation is used to calculate an upper limit
on the cosmogenic neutrino flux.

The effective target volume Fig.~\ref{fig:effvolume} was calculated
following the approach used for the first acoustic neutrino limit
estimate \cite{Vandenbroucke:2004gv}.  The neutrinos were assumed to
be down-going and to be uniformly distributed on a 2$\pi$ half sphere.
The total cross-section is taken from a function derived by extrapolation of
 measured cross sections to higher energies \cite{Ralston:1996bb}, which is
 valid for neutrino energies above $10^5$ GeV. 
Together with the interaction vertex, the direction $(\theta, \phi)$
defines the plane of the acoustic pressure wave perpendicular to this
 direction.  The sensor
observation angle was then calculated relative to this plane for each
vertex.  A number of $10^7$ events were simulated for neutrino
energies $E_{\nu}$ from $10^{18} \unit{eV}$ to $10^{22}
\unit{eV}$.  
The energy $E_\mathrm{had}$ of the hadronic cascade was assumed to be a constant
 fraction $y$ = 0.2 of the neutrino energy, i.e.~$E_\mathrm{had} = 0.2~E_{\nu}$.
The contribution of the cascade originating from the final state electron in
 the electron-neutrino charged current reaction is omitted in the present model
calculations, because its acoustic signal is expected to be small due to the
LPM-effect \cite{Niess:2005fw}. The acoustic pressure
$P_\mathrm{max}$ was calculated with respect to observation angle and distance.
We use the Askaryan \cite{Askaryan:1979vn} model to calculate the
acoustic signal strength assuming a cylindrical energy deposition in
the medium of length $L$ and diameter $d$.  No attempt was made to
model angular sensitivity or frequency response of the sensor.  A
minimum threshold of $\sim 300 \unit{mV}$, as in the real SPATS
measurement, was applied.  Using our estimate for the average SPATS
sensor sensitivity (Sec.~\ref{sec:absolute-noise}), this transforms to
a necessary minimum pressure of $\sim 70 \unit{mPa}$.  At least five
hits distributed over all four strings were required for an event to
trigger.

The procedure has a number of rather large associated uncertainties
that are discussed in the following list:

\begin{itemize}
\item The biggest uncertainty is related to the different predictions
  of the various thermo-acoustic models for the acoustic signal
  strength (see e.g. \cite{Butkevich:1998ic}, \cite{Bevan:2009kd}).
  We assume a 100 \% uncertainty on this quantity.  No absolutely
  calibrated measurement exists so far in any medium to fix that
  problem.
\item The Landau-Pomeranchuk-Migdal effect \cite{Migdal:1956} adds an
  additional uncertainty by changing the cross sections of
  bremsstrahlung and pair-production at ultra-high energies.
  The effect elongates preferentially
  primary electron cascades from neutrino interactions and diminishes
  expected acoustic signals \cite{Niess:2005fw}.  It starts to
  become important for hadronic cascades above $10^{18} \unit{eV}$.
  Above $10^{20} \unit{eV}$ its influence is
  diminished by photo-nuclear and electro-nuclear interactions (see
  \cite{Gerhardt:2010bj} for a recent detailed discussion).
\item The uncertainty of about 40\% in the absolute noise level
  determination leads to a corresponding uncertain trigger threshold.
  This makes the lower energy threshold for contributing neutrino
  interactions uncertain.
\item The angular efficiency loss of the single sensor channels is elaborated
 by use of a data set where two channels per sensor are available. 
With the threshold taken (70 mPa), an efficiency of $>$ 99\% is found for 99\%
 of the azimuthal angular range. 
This leads to the conclusion that the azimuthal efficiency loss is negligible
 in comparison to other error sources taken into account.  
\item The 30\% error on the sound attenuation length is of minor
  importance in comparison to the effects discussed above.
\end{itemize}

These problems influence all acoustic (and partly radio) neutrino
limits given so far (see Fig.~\ref{fig:sensitivity}).  In the present
paper the acoustic signal is calculated with a model delivering signal
values in the middle between extreme predictions.

The observation of zero events inside the effective volume of SPATS
gives an upper limit of $N_\mathrm{obs}=2.44$ events at a Poissonian
$90\%$ confidence level \cite{Feldman:1997qc}.  The flux limit for an
assumed trigger threshold of the measurement of 70 mPa is shown in
Fig.~\ref{fig:sensitivity} as dashed curve together with cosmogenic
neutrino flux predictions and results from other experiments. 
The grey and dark grey bands around the given limit indicate the
uncertainties of this quantity due to the uncertainties discussed
above. 
The upper border of the grey shaded area can therefore be considered as
 a conservative neutrino flux limit derived from the SPATS data, provided that
 the assumptions made in this work for estimating the detector sensitivity
 hold.

\section{Summary and Outlook}

We presented an analysis of acoustic noise data recorded with the
South Pole Acoustic Test Setup (SPATS) in the deep Antarctic ice at
the geographic South Pole. We found the absolute noise level to be
extremely stable over time. Its estimated magnitude of $14 \unit{mPa}$
in the deep ice, below $200 \unit{m}$, is comparable to the noise in
the deep sea when weather conditions are calm 
\cite{Aguilar:2010oq,Riccobene:2009zz,Aynutdinov:2009bs}.  Studies of transient
noise in the SPATS data revealed the refreezing IceCube holes and
Rodriguez-Wells as sources. The high quality of the data allowed us to
study the refreezing processes as function of time in great detail. No
transient acoustic signals in the deep ice were observed outside the
instrumented volume of IceCube at depths below $200 \unit{m}$.  This
enabled us to derive a first upper limit on the flux of ultra-high
energy neutrinos with an acoustic detector in glacial ice.

SPATS is continuing to take data. An upgrade of the DAQ software to
read out all sensor channels simultaneously and to form a multiplicity
trigger online, will increase the detector sensitivity.

\section*{Acknowledgements}

We acknowledge the support from the following agencies: U.S.~National
Science Foundation-Office of Polar Programs, U.S.~National Science
Foundation-Physics Division, University of Wisconsin Alumni Research
Foundation, the Grid Laboratory Of Wisconsin (GLOW) grid
infrastructure at the University of Wisconsin -- Madison, the Open
Science Grid (OSG) grid infrastructure; U.S.~Department of Energy, and
National Energy Research Scientific Computing Center, the Louisiana
Optical Network Initiative (LONI) grid computing resources; National
Science and Engineering Research Council of Canada; Swedish Research
Council, Swedish Polar Research Secretariat, Swedish National
Infrastructure for Computing (SNIC), and Knut and Alice Wallenberg
Foundation, Sweden; German Ministry for Education and Research (BMBF),
Deutsche Forschungsgemeinschaft (DFG), Research Department of Plasmas
with Complex Interactions (Bochum), Germany; Fund for Scientific
Research (FNRS-FWO), FWO Odysseus programme, Flanders Institute to
encourage scientific and technological research in industry (IWT),
Belgian Federal Science Policy Office (Belspo); University of Oxford,
United Kingdom; Marsden Fund, New Zealand; Japan Society for Promotion
of Science (JSPS); the Swiss National Science Foundation (SNSF),
Switzerland; A.~Gro{\ss} acknowledges support by the EU Marie Curie
OIF Program; J.~P.~Rodrigues acknowledges support by the Capes
Foundation, Ministry of Education of Brazil.


\end{document}